\documentclass[fleqn,usenatbib]{mnras}

\usepackage{newtxtext,newtxmath}

\usepackage[T1]{fontenc}
\usepackage{ae,aecompl}

\usepackage{graphicx}	
\usepackage{amsmath}	
\usepackage{amssymb}	
\usepackage[english]{babel}
\usepackage[utf8]{inputenc}
\usepackage[colorinlistoftodos]{todonotes}
\usepackage{xcolor}
\usepackage{ulem}
\usepackage{censor}
\usepackage{pdflscape}

\newcommand{\galex}{\textit{GALEX}}
\newcommand{\galexfull}{\textit{Galaxy Evolution Explorer}}
\newcommand{\sdss}{\textit{SDSS}}
\newcommand{\sdssfull}{\textit{Sloan Digital Sky Survey}}
\newif\ifrev
\revfalse
\newif\ifdel
\delfalse
\newlength\nextcharwidth
\makeatletter
\renewcommand\@cenword[1]{%
  \setlength{\nextcharwidth}{\widthof{#1}}%
  \censorrule{\nextcharwidth}%
  \kern -\nextcharwidth%
  #1}
\makeatother
\newcommand{\delref}[1]{\ifdel\textcolor{brown}{\censorruledepth=.55ex\censor{#1}}\else\fi}

\newcommand{\rev}[1]{\ifrev\textcolor{brown}{#1}\else{#1}\fi}
\newcommand{\rrev}[1]{\ifrev\textcolor{brown}{\textbf{#1}}\else{#1}\fi}
\newcommand{\del}[1]{\ifdel\textcolor{brown}{\sout{#1}}\else\fi}
\newcommand{\nrev}[1]{\ifrev\textcolor{magenta}{#1}\else{#1}\fi}
\newcommand{\nrrev}[1]{\ifrev\textcolor{magenta}{\textbf{#1}}\else{#1}\fi}
\newcommand{\ndel}[1]{\ifdel\textcolor{red}{\sout{#1}}\else\fi}
\newcommand{\nddel}[1]{\ifdel\textcolor{red}{\sout{\textbf{#1}}}\else\fi}

\newcommand{\kms}{km s$^{-1}$}

\title[\nrev{Determination of RT Model Parameter Uncertainties}\ndel{Revisiting EON\_10.477\_41.954 (FGC 79)}]{\nrev{Seeking Edge-on Galaxies with Substantial Extraplanar Dust Using a Radiative Transfer Model: Determination of the Model Parameter Uncertainties for EON\_10.477\_41.954 (FGC 79)}\ndel{Revisiting EON\_10.477\_41.954 (FGC 79), a Target from a Candidate List of Edge-on Galaxies with Substantial Extraplanar Dust}}

\author[J.-H. Shinn]{
Jong-Ho Shinn,$^{1}$\thanks{E-mail: jhshinn@kasi.re.kr}
\\
$^{1}$Korea Astronomy and Space Science Institute, 776 Daeduk-daero, Yuseong-gu, Daejeon, 305-348, the Republic of Korea
}

\date{Accepted XXX. Received YYY; in original form ZZZ}

\pubyear{2019}

\hypersetup{draft}
\begin{document}
\label{firstpage}
\pagerange{\pageref{firstpage}--\pageref{lastpage}}
\maketitle

\begin{abstract}
We have revisited the target EON\_10.477\_41.954 in order to determine more accurately the uncertainties in the model parameters that are important for target classification \rev{(i.e., galaxies with or without substantial extraplanar dust)}.
We performed a Markov-Chain Monte Carlo (MCMC) analysis for the fifteen parameters of the three-dimensional radiative-transfer galaxy model we used previously for target classification.
To investigate the convergence of the MCMC sampling\del{, }\rev{---which is usually neglected in the literature but should not be---}we monitored the integrated autocorrelation time ($\tau_{int}$), and we achieved effective sample sizes $>5,650$ for all the model parameters.
The confidence intervals are unstable at the beginning of the iterations where the values of $\tau_{int}$ are increasing, but they become stable in later iterations where those values are almost constant.
The final confidence intervals are \nddel{not equal to}\ndel{---and not even constant multiples of---}\nddel{---but $\sim10-100$ times larger than---}\nrrev{$\sim5-100$ times larger than} the nominal uncertainties used in our previous study (the standard deviation of three best-fit results).
Thus, those nominal uncertainties are not good proxies for the model-parameter uncertainties.
\nddel{However, although}\nrrev{Although} the position of EON\_10.477\_41.954 in the target-classification plot \rev{(the scale-height to diameter ratio of dust vs that of light-source)} decreases by about $20-30$\% when compared to our previous study, its membership in the ``high-group''---i.e., among galaxies with substantial extraplanar dust---nevertheless remains unchanged.
\end{abstract}




\section{Introduction \label{intro}}
Galactic dust \nrev{which resides in various types of galaxy (spiral, elliptical and irregular)} manifests itself by scattering, absorbing, or reradiating starlight \citep{Stein_1983_ARA&A_21_177}.
\nrev{The optical dust lane seen along the galactic plane of the spiral galaxy is a clear manifestation of its existence.}
The dust located away from \ndel{the}\nrev{this} galactic plane (i.e., extraplanar dust) is important in several respects.
For instance, the very presence of extraplanar dust indicates that some mechanism prevents the dust from settling down to the galactic plane.
\nrev{\cite{Howk_1997_AJ_114_2463} discussed several mechanisms that might produce the extraplanar dust, reporting its first clear detection. They include hydrodynamic phenomena such as galactic fountain flows \citep{Shapiro_1976_ApJ_205_762,Bregman_1980_ApJ_236_577,Houck_1990_ApJ_352_506}, radiation pressure \citep{Ferrara_1991_ApJ_381_137,Franco_1991_ApJ_366_443,Ferrara_1993_ApJ_407_157}, effects of magnetic fields and dynamical instabilities \citep{Binney_1981_MNRAS_196_455,Mulder_1984_A&A_134_158}. Even the accretion from the circumgalactic or intergalactic medium (see \citealt{Howk_2012_inproc} and \citealt{Fraternali_2008_MNRAS_386_935}) and the tidal or ram-pressure stripping from galaxies \citep{Hodges-Kluck_2019_inproc} are on the list.}
Extraplanar dust may also contain information about the transport of the interstellar medium into the galactic halo and out into intergalactic space\nrev{, where the dust is known to exist \citep[e.g.,][]{Zaritsky_1994_ApJ_435_599,Chelouche_2007_ApJ_671_L97,McGee_2010_MNRAS_405_2069,Menard_2010_MNRAS_405_1025,Peek_2015_ApJ_813_7,Tumlinson_2017_ARA&A_55_389}}.

Numerous studies have been done on extraplanar dust at several wavelengths, mainly targeting edge-on galaxies.
Its existence is indicated by optical absorption features detected against starlight \citep[e.g.,][]{Howk_1997_AJ_114_2463,Howk_1999_AJ_117_2077,Howk_1999_Ap&SS_269_293,Thompson_2004_AJ_128_662,Rossa_2004_AJ_128_674}, by emission features in the infrared \citep[e.g.,][]{Irwin_2006_A&A_445_123,Irwin_2007_A&A_474_461,Burgdorf_2007_ApJ_668_918,Bocchio_2016_A&A_586_A8}, and by scattered light in the ultraviolet \citep[e.g.,][]{Seon_2014_ApJ_785_L18,Hodges-Kluck_2014_ApJ_789_131,Shinn_2015_ApJ_815_133,Hodges-Kluck_2016_ApJ_833_58}.
Extraplanar dust has also been studied by modeling galactic spectral-energy distributions \citep{Baes_2016_A&A_587_A86} and by modeling the polarization of starlight \citep{Seon_2018_ApJ_862_87}.
The typical scale-height of extraplanar dust is a few kpc.
Two investigations \citep{Hodges-Kluck_2014_ApJ_789_131,Hodges-Kluck_2016_ApJ_833_58} studied the relations between the properties of ultraviolet halos and their host galaxies, but \cite{Hodges-Kluck_2016_ApJ_833_58} found no close relation between the scale-height of the ultraviolet halo and host-galaxy properties such as the galaxy luminosity or the star-formation rate.
Note, however, that the scale-height studied by \cite{Hodges-Kluck_2016_ApJ_833_58} is for the dust-scattered ultraviolet halo and not for the dust itself.

One way to characterize extraplanar dust in the context of galaxy evolution is to examine its relation to the properties of the host galaxies.
In order to derive the physical parameters of the extraplanar dust (e.g., the scale-height) and of the host galaxies (e.g., the star-formation rate), we previously examined those galaxies reported to have dust-scattered ultraviolet halos \citep{Hodges-Kluck_2014_ApJ_789_131}, using a three-dimensional radiative-transfer model \citep{Shinn_2015_ApJ_815_133}.
However, our results showed that the existence of an ultraviolet halo does not always mean the existence of substantial extraplanar dust.
Therefore, we needed a more reliable list of galaxies with substantial extraplanar dust, and we constructed a new candidate list by examining edge-on galaxies \citep{Shinn_2018_ApJS_239_21}.
In that list, the candidates are those for which the stellar and dust scale-heights ($z_s$ and $z_d$) are hard to explain with a simple disk model (one stellar disk + one dust disk)\rev{, probably due to the existence of extraplanar dust}.

We obtained these scale-heights $z_s$ and $z_d$, in addition to thirteen other model parameters, by fitting ultraviolet galaxy images with a three-dimensional radiative-transfer model \citep{Shinn_2018_ApJS_239_21}.
For the nominal uncertainties of the model parameters, in that study we used the standard deviation of three best-fit values.
We obtained those values by repeating the fit three times, similarly as done by \cite{DeGeyter_2013_A&A_550_A74,DeGeyter_2014_MNRAS_441_869}.
However, the resulting nominal uncertainties are \del{only }rough estimates.
Because they may affect target classification in the plane $z_d/D_{25,ph}$ vs. $z_s/D_{25,ph}$, where $D_{25,ph}$ is the galactic diameter \citep[see][]{Shinn_2018_ApJS_239_21}, in the present work we have revisited one target---EON\_10.477\_41.954 (FGC 79)---and have performed a Markov-Chain Monte Carlo (MCMC) analysis (see \citealt{Sharma_2017_ARA&A_55_213} and references therein) to check this possibility and obtain more rigorous estimates of the model-parameter uncertainties.
\rev{While applying the MCMC method, we monitored the convergence of the MCMC samples---which is usually neglected in the literature but should not be---in order to know how much the samples are close to the targeted distribution, i.e., the likelihood (see section \ref{ana-res}).}
The well-sampled MCMC analysis shows that the new parameter estimates are consistent with the previous ones, in general to within the 95\% confidence interval.
However, we also found the nominal uncertainties used in \cite{Shinn_2018_ApJS_239_21} to be unsatisfactory, since their ratios to the MCMC confidence intervals are far from unity and are not even constant.
Nevertheless, we have also found that the target classification of EON\_10.477\_41.954 in the plane $z_d/D_{25,ph}$ vs. $z_s/D_{25,ph}$ remains unchanged.

\section{Data and Target \label{obs-red}} 
Since we are here revisiting one target from among the edge-on galaxies studied by \cite{Shinn_2018_ApJS_239_21}, our data-preparation procedure is exactly the same as in that paper.
We used the FUV-band image from the \galexfull{} (\galex{}; \citealt{Martin_2005_ApJ_619_L1}), which covers 1344–1786 \AA{} with an imaging resolution of $4.2''$ \citep{Morrissey_2007_ApJS_173_682}.
\rev{The FUV-band image is useful in discovering extraplanar dust because the dominant ultraviolet light sources---massive stars---have smaller scale-height than the old stellar population \citep{Bahcall_1980_ApJS_44_73,Wainscoat_1992_ApJS_83_111,Martig_2014_MNRAS_442_2474}.}
We used the pipeline-processed archival data from GR6/7 (\url{http://galex.stsci.edu/GR6/}) without any additional data reduction except for cropping and masking.
A more detailed description of the data preparation can be found in \cite{Shinn_2018_ApJS_239_21}.

\begin{figure*}
\center{
\includegraphics[scale=0.8]{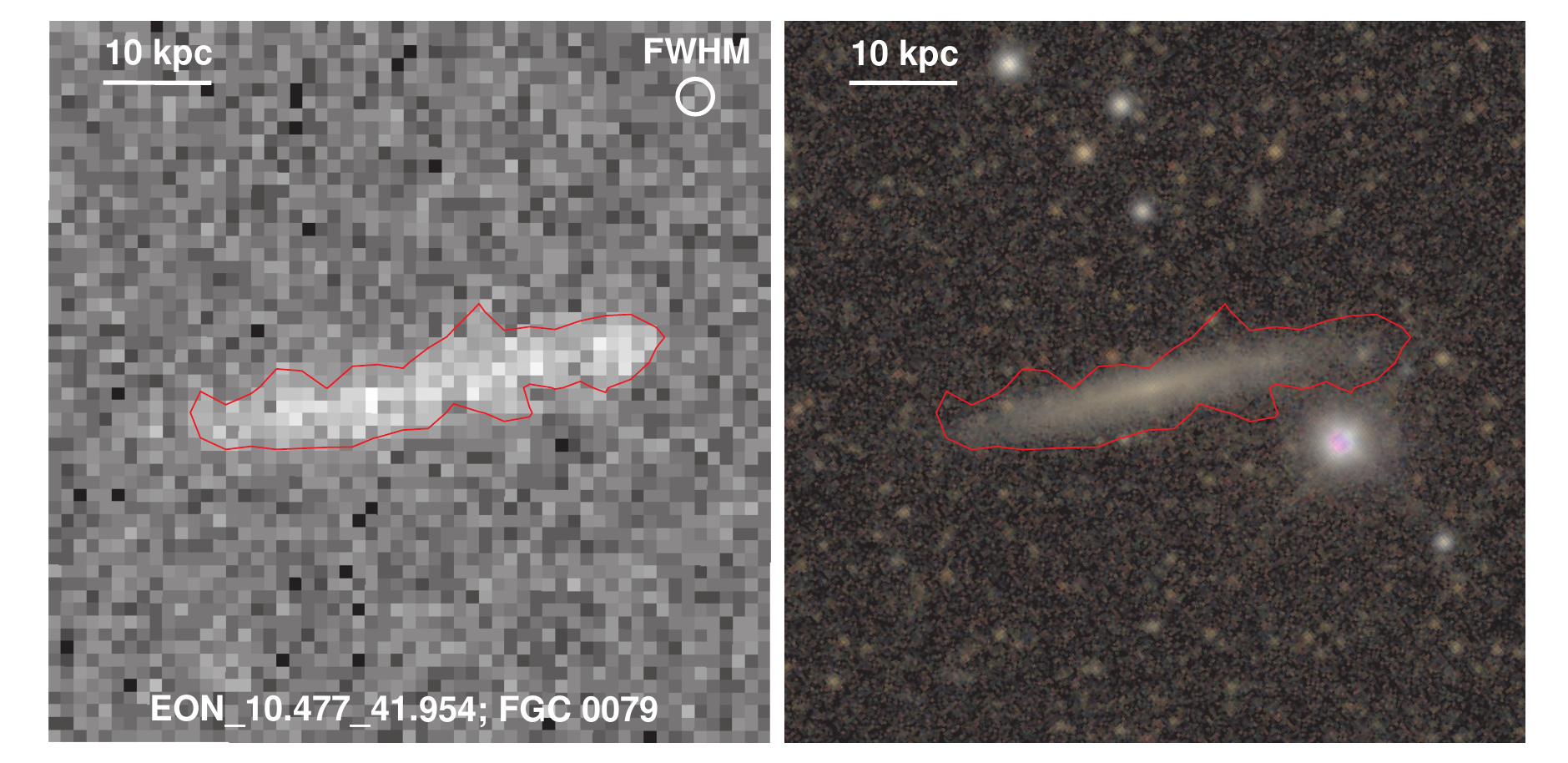}
}
\caption{\galexfull{} (\galex) FUV image (\textit{left}) and \sdssfull{} (\sdss, \protect\citealt{York_2000_AJ_120_1579}) RGB image (\textit{right}) of EON\_10.477\_41.954.
\rev{Two images are in the same orientation and scale.}
\nrev{The red lines indicate the 3-$\sigma$ level determined from the \galex{} FUV count image. The circle at the upper-right corner in the left panel is the imaging resolution of the \galex{} FUV band ($4.2''$, \citealt{Morrissey_2007_ApJS_173_682}).}
The \sdss{} RGB image is made from $g$=blue, $r$=green, and $i$=red.
} \label{fig-tg}
\end{figure*}

We selected the target EON\_10.477\_41.954 for this study because \rrev{it seems to have a simple unimodal likelihood surface which is more tractable for the MCMC analysis---}its FUV image has \rrev{a smooth overall shape and }few local (bright or dark) features\rrev{,} and was well-fitted by the model, leaving an almost featureless residual image, as shown in \cite{Shinn_2018_ApJS_239_21}.
Figure \ref{fig-tg} shows images of EON\_10.477\_41.954 in the far-ultraviolet (the left panel) and the optical (the right panel), respectively.

\section{Analysis and Results} \label{ana-res}
The primary purpose of this study is to determine the uncertainties in the model parameters more accurately than we did in \cite{Shinn_2018_ApJS_239_21}.
In order to accomplish this, we employed the MCMC method and sampled the likelihood surface.
The three-dimensional radiative-transfer model and the likelihood calculation are the same as in \cite{Shinn_2018_ApJS_239_21}.
The model produces a two-dimensional image at an observer's position for a given set of parameters, repeatedly generating an artificial distribution of photons from the galactic light source and scattering them from the galactic dust within the three-dimensional model grid.

We used fifteen parameters for the modeling, eight for the galaxy itself ($L_{FUV},\,\tau_{FUV},\,h_d,\,z_d,\,R_d,\,h_s,\,z_s,\,R_s$) and seven for the viewpoint and the background ($\theta_{incl},\,\theta_{pos},\,\Delta x,\,\Delta y,\,bg_{ctr},\,\partial bg/\partial x,\,\partial bg/\partial y$).
\rev{The galaxy is described by two exponential discs, one for the dust and the other for the light-source.}
The parameter definitions are as follows: $L_{FUV}$ is the galaxy luminosity in the FUV band; $\tau_{FUV}$ is the optical depth in the FUV band \nrev{through the galactic center when viewed face-on}; $h_d$ is the scale-length of the dust component; $z_d$ is the scale-height of the dust component; $R_d$ is the cutoff radius of the dust component; $h_s$ is the scale-length of the light-source component; $z_s$ is the scale-height of the light-source component; $R_s$ is the cutoff radius of the light-source component; $\theta_{incl}$ is the inclination angle; $\theta_{pos}$ is the position angle; $\Delta x$ is the horizontal shift of the galaxy center in the image plane; $\Delta y$ is the vertical shift of the galaxy center in the image plane; $bg_{ctr}$ is the background level at the center of the image; $\partial bg/\partial x$ is the horizontal gradient of the background at the center of the image; and $\partial bg/\partial y$ is the vertical gradient of the background at the center of the image.
\nrev{The viewpoint and the background are included in the model, to obtain more general uncertainties of galaxy parameters which reflect the uncertainties of the viewpoint and the background.}
We refer the readers to \cite{Shinn_2018_ApJS_239_21} for more details about the model.

We calculated the likelihood using the $C$-statistic \citep{Cash_1979_ApJ_228_939} in order to reflect the low photon counts in the \galex{} FUV image: $C=-2\,\Sigma_{i}(d_i \ln m_i - m_i - \ln d_i !)$, where $m_i$ and $d_i$ are the pixel values of the model and data images, respectively.
Before calculating the likelihood, we convolved the model image with the extended point-spread-function (an image size of $600''$) to match the spatial resolution and include instrumental light scattering (see \citealt{Shinn_2018_ApJS_239_21} for more details).
\rev{Here we note that the vertical profile of the dust-scattered starlight can have a broader width than that of the dust itself, since the starlight can move away from the galactic plane being scattered multiple times by the dust.}

\subsection{MCMC Sampling} \label{ana-res-mcmc}
We adopted the affine-invariant ensemble sampler \citep{Goodman_2010_CAMCoS_5_65} for the MCMC sampling.
This sampling method uses a set of walkers that move around the target distribution, changing their positions according to the affine transformation.
We used the \textit{stretch move} method, which generates a proposal from the positions of the current walker and one other randomly selected walker.
This can be expressed as follows:
\begin{equation}
    X_k \xrightarrow{} Y = X_j + Z\times(X_k(t) - X_j)
\end{equation}
where $X_j$ and $X_k$ represent individual walkers in the ensemble, $Y$ is a proposal for the walker $X_k$, and $Z$ is a scaling variable.
The distribution of $Z$ is given by
\begin{equation} \label{eq-gz}
    g(z) \propto
    \begin{cases}
        \dfrac{1}{\sqrt{z}} & \text{if $z \in \left[ \dfrac{1}{a},a \right] $ } \\
        0 & \text{otherwise}
    \end{cases}
\end{equation}
where the parameter $a>1$ is adjustable.
We adopted $a=2$, as in \cite{Goodman_2010_CAMCoS_5_65}.
In summary, our MCMC sampler makes a proposal for $X_k$ along the line that connects $X_k$ and $X_j$, being limited by the two points [$X_j + \frac{1}{2}\times(X_k - X_j)$] and [$X_j + 2\times(X_k - X_j)$], which is biased more toward $X_j$ than in the opposite direction [see eq.~(\ref{eq-gz})].

To save time in finding the maximum likelihood region, we used as the initial MCMC walkers the model-parameter vectors obtained by \cite{Shinn_2018_ApJS_239_21} after finding the best fit.
The number of walkers is hence 113, which was the number of parameter vectors used in \cite{Shinn_2018_ApJS_239_21}.
The MCMC walkers seem to be in the ``burn-in'' stage over about the first 4000 iterations \citep[see][for more about ``burn-in'']{Brooks_2011_book}.
We excluded the MCMC chains that were produced during this ``burn-in'' stage.

To perform the MCMC analysis, we utilized a computer cluster, running the calculation either with 256 CPUs or with 240 CPUs.
The CPU models employed for the former and the latter were, respectively, Intel(R) Xeon(R) CPU E5-2470 0 @ 2.30GHz and Intel(R) Xeon(R) CPU E5-2698 v4 @ 2.20GHz.
The running time was almost the same for both, requiring about five hours for 1000 iterations, or equivalently about one day for 5000 iterations.
We produced each model image with $5\times10^6$ photons, as in \cite{Shinn_2018_ApJS_239_21}.
We note that the computing speed is about three times faster than we had available for \cite{Shinn_2018_ApJS_239_21}.

\subsection{Convergence Monitoring} \label{ana-res-cnvg}
The MCMC algorithm is a method for sampling the target distribution, and it is important to check whether the samples are sufficiently representative of that distribution; in other words, whether the samples are adequately converged.
\rev{This convergence monitoring should be carried out; if not, we cannot have any information about how much the samples are close to, or different from, the target distribution.}
The degree of convergence is directly related to the reliability of the model-parameter uncertainties, since we are sampling the likelihood surface.
In order to monitor the convergence of the MCMC chains, we measured the integrated autocorrelation time ($\tau_{int}$\nrev{, see \citealt{Sharma_2017_ARA&A_55_213,Sokal_2013_inbook}}), which is defined as
\begin{eqnarray} \label{eq-iat}
    \tau_{int}=\sum^{\infty}_{t=-\infty}\rho_{xx}(t)
    \text{, where } \rho_{xx}(t)=\frac{\mathbb{E}[(x_i-\Bar{x})(x_{i+t}-\Bar{x})]}{\mathbb{E}[(x_i-\Bar{x})^2]}
\end{eqnarray}
Here $\rho_{xx}$ is the autocorrelation function for a sequence $\{x_i\}$, $t$ is the time difference---or distance---between two points in the sequence $\{x_i\}$, $\Bar{x}$ is the mean of sequence $\{x_i\}$, and $\mathbb{E}\left[\cdot\right]$ means the expectation value.
Note that $\rho_{xx}$ has a form similar to Pearson's correlation coefficient \nrev{(see \citealt{Press_2007_book})}.

In most applications, the MCMC chains are correlated \citep[see][]{Sokal_2013_inbook}, and the variance of $\Bar{x}$ increases by the factor $\tau_{int}$:
\begin{equation} \label{eq-var}
    \text{Var}(\Bar{x})\simeq\tau_{int}\,\frac{\mathbb{E}[(x_i-\Bar{x})^2]}{N}
\end{equation}
where $N$ is the size of the sequence $\{x_i\}$ \citep{Sharma_2017_ARA&A_55_213,Sokal_2013_inbook}.
In other words, we can say that the MCMC chain has only $N/\tau_{int}$ independent samples, which is also known as ``effective sample size'' (ESS).
Based on Eq.~(\ref{eq-var}), we can monitor the convergence of the MCMC chain using the ESS, since Var($\Bar{x}$) becomes smaller as the MCMC samples approach the target distribution more closely.

\begin{figure*}
\center{
\includegraphics[scale=0.45]{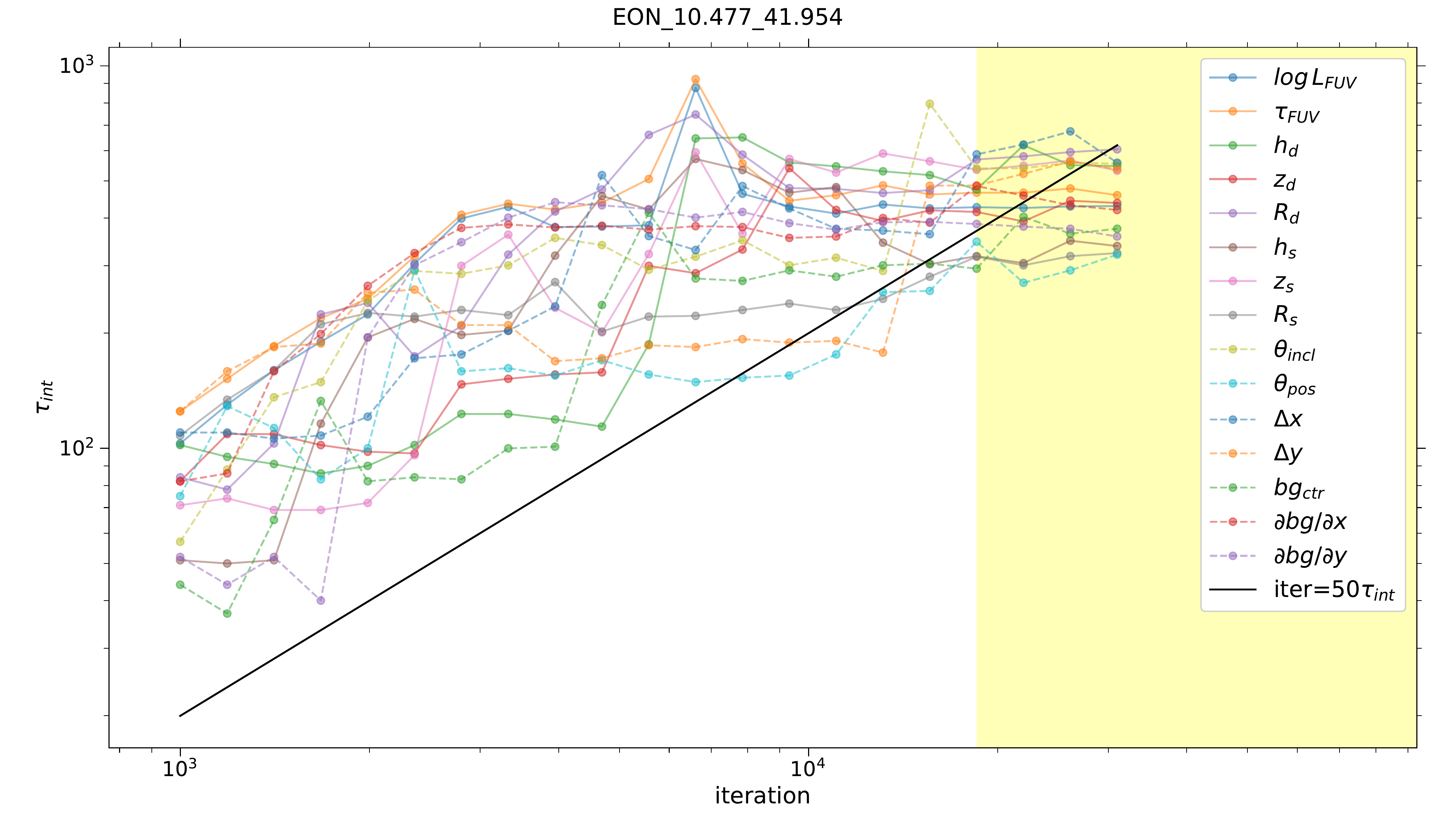}
}
\caption{Evolution of the integrated autocorrelation time $\tau_{int}$ \nrev{for EON\_10.477\_41.954}. Each colored line corresponds to the model parameter listed in the legend. The yellow shaded region shows where all the Markov-Chain Monte Carlo (MCMC) chains have almost constant values of $\tau_{int}$. The black solid line indicates the locus of iteration$=50\times\tau_{int}$.} \label{fig-iat}
\end{figure*}

Fig.~\ref{fig-iat} shows the values of $\tau_{int}$ for each of our model parameters.
Since we used the affine-invariant ensemble sampler, we calculated $\tau_{int}$ from the mean of the 113 walkers $\{X_i\}$, where $X_i=\frac{1}{113}\sum^{113}_{i=1} x_i$, as noted in \cite{Goodman_2010_CAMCoS_5_65}.
We calculated $\tau_{int}$ with a routine in the MCMC package \textsf{emcee} \citep{Foreman-Mackey_2013_PASP_125_306}, using an ``automatic windowing'' size of 5 \citep[see][]{Sokal_2013_inbook}.
At the beginning of the iterations, the values of $\tau_{int}$ increase with the number of the iterations, but they saturate at the level of $\sim300-600$ after $\sim2\times10^4$ iterations.
This saturated value can be regarded as the representative value of $\tau_{int}$ for our MCMC method.
As shown in Fig.~\ref{fig-iat}, the values of $\tau_{int}$ for all the parameters cross the line $N=50\,\tau_{int}$ at the end of the iterations.
This means that all the MCMC chains have ESSs greater than $50\times113=$ 5,650; the number of walkers, 113, must be multiplied to, since the value of $\tau_{int}$ is calculated from the mean of ensemble walkers ($X_i=\frac{1}{113}\sum^{113}_{i=1} x_i$, see also \citealt{Goodman_2010_CAMCoS_5_65}).
This ESS of 5,650 far exceeds the ESS $=3,748$ required to determine the 0.025 quantile of \ndel{$\Bar{x}$ [see eq.~(\ref{eq-var})]}\nrev{the probability distribution function $f(x)$} to within $\pm0.005$ of the true value with probability 0.95 \citep{Raftery_1992_inproc}.
We therefore conclude that our MCMC chains achieve sufficiently large ESSs and hence good enough convergence.

We expected the running time for all 35,000 iterations (including the ``burn-in'' stage; see section \ref{ana-res-mcmc}) to be about one week, but we actually had to spend much more time.
At some points the MCMC chains tended to become similar to earlier ones, which increases $\rho_{xx}(t)$ and the values of $\tau_{int}$ but without increasing the ESS.
We therefore had to stop at such points and resume the MCMC sampling after removing the last parts of those MCMC chains.
This happened several times before we completed all 35,000 iterations.

When carrying out the MCMC analysis, we used as the initial MCMC walkers the model parameter vectors obtained by \cite{Shinn_2018_ApJS_239_21} after finding the best fit (see section \ref{ana-res-mcmc}).
This approach might miss other local maxima of the likelihood surface, located away from the global maximum.
We therefore checked for such possibilities by running a test MCMC sampling with a larger proposal distance, increasing the parameter $a$ in Eq.~(\ref{eq-gz}) five-fold from 2 to 10.
We adopted the last 113 walkers---that is, those obtained at the 35,000th iteration---as the initial MCMC walkers for this test, and we ran the sampling for up to 300 iterations.
We found no local maxima from this test.

\subsection{Confidence Intervals from MCMC Sampling}
\begin{figure*}
\center{
\includegraphics[scale=0.45]{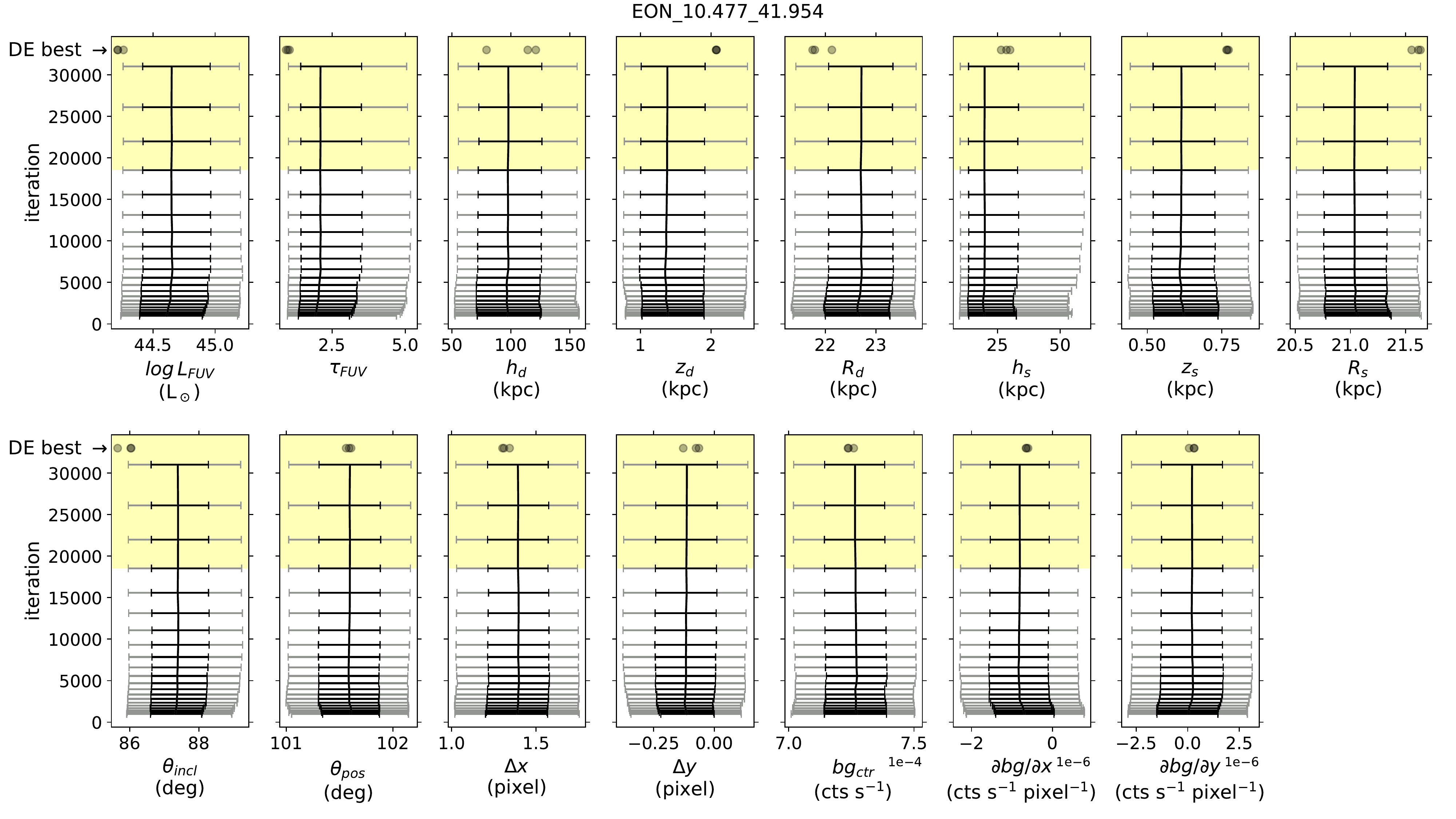}
}
\caption{Evolution of the confidence intervals \nrev{for EON\_10.477\_41.954}. The vertical black line in each panel represents the locus of the median value. The black and gray error bars are the 68\% and 95\% confidence intervals, respectively. The yellow shaded region is the same as in Fig.~\ref{fig-iat}, where all the Markov-Chain Monte Carlo (MCMC) chains show almost constant values of $\tau_{int}$. The three gray circles at the top of each panel represent the best-fit results obtained in \protect\cite{Shinn_2018_ApJS_239_21} using the Differential Evolution (DE) algorithm; hence we label them ``DE best.''} \label{fig-conf_evol}
\end{figure*}

The purpose of our MCMC analysis is to determine the uncertainties of our model parameters more accurately, and the accuracy depends upon the convergence of the MCMC sampling.
In order to see this dependence, we plotted the evolution of the uncertainties during the iterations.
Fig.~\ref{fig-conf_evol} shows the evolution of the 68\% and 95\% confidence intervals.
Both confidence intervals vary in position and width at the beginning of the iterations, but they stabilize later on, especially after entering the region where the values of $\tau_{int}$ become almost constant (the yellow shaded region in Fig.~\ref{fig-conf_evol}).
The ``burn-in'' stage is not included in Fig.~\ref{fig-conf_evol} (see section \ref{ana-res-mcmc}).
We can see the convergence of the MCMC sampling from Fig.~\ref{fig-conf_evol}, although the converged confidence intervals are not greatly different from the earlier ones.

For comparison with the best-fit model parameters obtained by \cite{Shinn_2018_ApJS_239_21}, we plotted those best-fit values in Fig.~\ref{fig-conf_evol} with the label ``DE best.''
\cite{Shinn_2018_ApJS_239_21} found these best-fit values through a global-optimization method called Differential Evolution (DE).
The three ``DE best'' points correspond to three different results of DE optimization.
As shown in Fig.~\ref{fig-conf_evol}, the ``DE best'' points fall outside the 68\% confidence interval for some model parameters, such as $L_{FUV}$ and $\tau_{FUV}$.
However, they generally seem to fall within the 95\% confidence interval.
\rev{We here note that the number of iteration required for the MCMC analysis ($35,000$) far exceeds that of the DE method ($\sim3000$ for three trials). This means that the MCMC analysis is much more time-consuming than the DE method.}

\begin{figure*}
\center{
\includegraphics[scale=0.2]{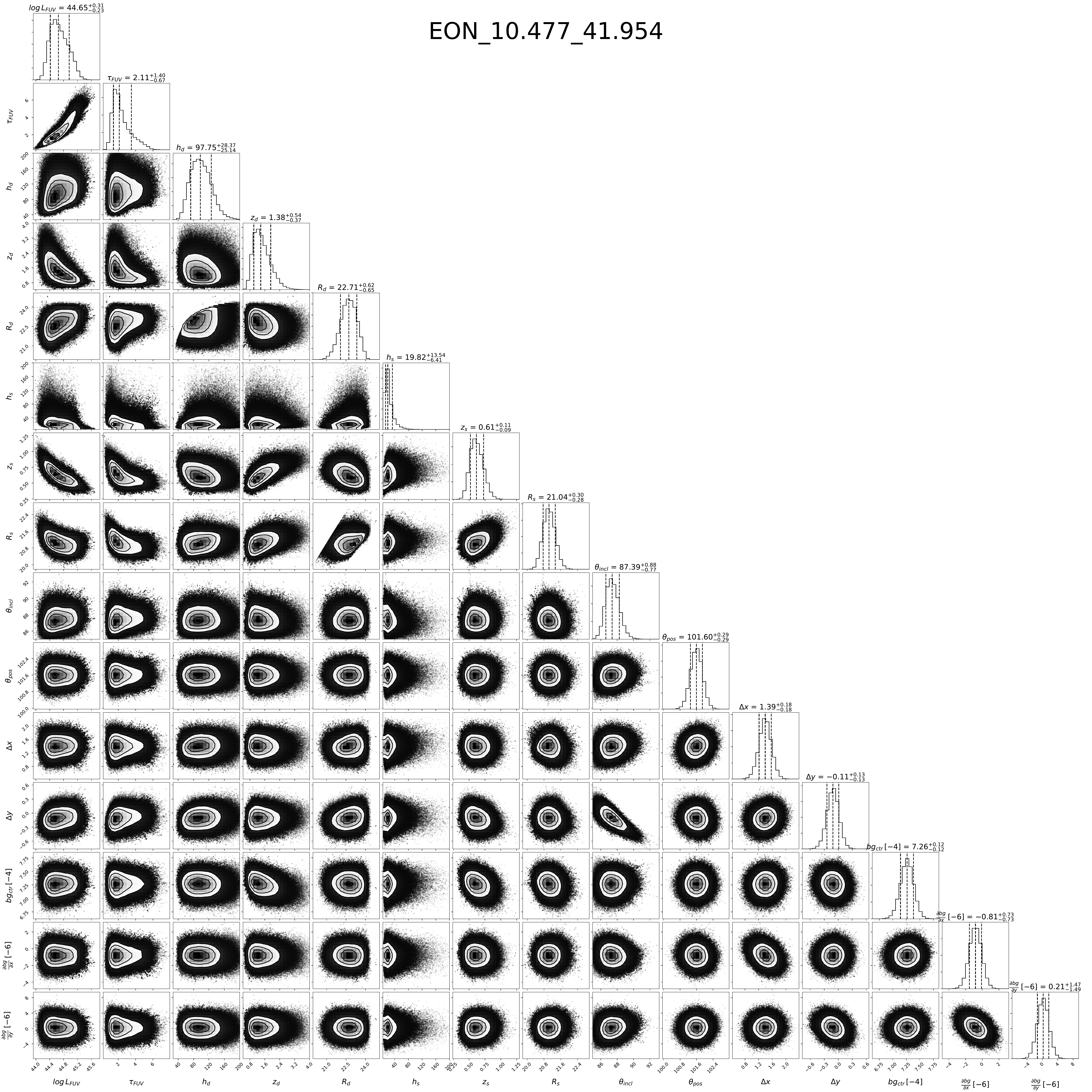}
}
\caption{Corner plot \ndel{of the 15 model parameters}\nrev{for EON\_10.477\_41.954}. The diagonal panels show the marginal probability distribution functions of the model parameters, where the three vertical dashed lines represent the median and 68\% confidence intervals. The off-diagonal panels show the pairwise relations, with no smoothing applied. \nrev{The lower end of the $h_s$ distribution is limited by the parameter space limit, 0.1 kpc. (See the separate sheet for a larger version of this figure.)}} \label{fig-corner}
\end{figure*}

Fig.~\ref{fig-corner} is a so-called ``corner plot\del{,''}\rev{'',} which shows all the mutual relations among the model parameters (the off-diagonal panels) and the marginal probability distribution functions (PDFs) for each parameter (the diagonal panels).
We created this corner plot from all the MCMC chains obtained over the final 31,000 iterations, i.e., after excluding the first 4,000 (see Fig.~\ref{fig-conf_evol}).
No smoothing was applied to the off-diagonal panels.
There is a sharp cut in the panel for $R_s$ vs. $R_d$, which is caused by the restriction of $R_d \geq R_s$; we used this constraint to prevent light sources from being in the space without dust (see \citealt{Shinn_2018_ApJS_239_21}).
There are clear correlations between some parameters.
For example, there is a correlation between $L_{FUV}$ and $\tau_{FUV}$, and there are anti-correlations between $L_{FUV}$ and $z_{s}$ and between $\Delta y$ and $\theta_{incl}$.
The correlation between $L_{FUV}$ and $\tau_{FUV}$ occurs because an increase in $L_{FUV}$ increases the brightness of the model image, while $\tau_{FUV}$ decreases it.
The anti-correlations between $L_{FUV}$ and $z_{s}$ can be understood similarly, because both parameters increase the brightness of the model image at the edge of the galaxy.
The anti-correlation between $\Delta y$ and $\theta_{incl}$ is related to the geometry of the galactic disk relative to the observer.
Since $\theta_{incl}$ is less than 90\degr, the upper disk surface faces the observer, so the projected disk area decreases as $\theta_{incl}$ increases toward 90\degr.
The brightness in the lower part of the galaxy-model image, which corresponds to the front side of the disk surface, is more affected by the galaxy inclination, since direct (non-scattered) light is more dominant there.
When $\theta_{incl}$ increases, $\Delta y$ must decrease to compensate for the darkening of the lower part of the galaxy-model image, and vice versa; this leads to the anti-correlation.
\rev{We here note the weak anti-correlation between $z_d$ and $h_d$. The scattering effect of extraplanar dust can be reproduced by increasing either $z_d$ or $h_d$, since our target galaxy is viewed edge-on. This degeneracy seems to cause the large $h_d\sim100$ kpc although its uncertainty is also large ($\sim20-30$ kpc).}
\rrev{We also observed that a substantial extraplanar dust causes the overestimation of $h_d$, from Test C (Fig.~4) of \cite{Shinn_2018_ApJS_239_21}.}
\nrrev{The parameter $h_s$ is limited at its lower end by the parameter space limit (0.1  kpc) that was set to be small enough considering the spatial resolution of the EON\_10.477\_41.954 FUV image.}
The marginal PDFs are shown in the diagonal panels of Fig.~\ref{fig-corner}, together with their medians and 68\% confidence intervals (see also Table \ref{tbl-bestfit}).
While a few parameters (such as $L_{FUV}$, $\tau_{FUV}$, and $z_d$) show skewed PDFs, the others show roughly symmetric PDFs.

\begin{landscape}
\begin{table}
    \caption{Best Model Parameters \nrev{for EON\_10.477\_41.954}. \rev{The results of \protect\cite{Shinn_2018_ApJS_239_21} are appended at the last row for comparison.} Column (1), the galaxy luminosity in the FUV band; Column (2), the optical depth in the FUV band \nrev{through the galactic center when viewed face-on}; Column (3), the scale-length of the dust component; Column (4), the scale-height of the dust component; Column (5), the cutoff radius of the dust component; Column (6), the scale-length of the light-source component; Column (7), the scale-height of the light-source component; Column (8), the cutoff radius of the light-source component; \del{and }Column (9), the inclination angle\rev{; and Column (10), a note for the row}. The quoted values \rev{of this work} are the medians and the 68\% confidence intervals\rev{, while those of \protect\cite{Shinn_2018_ApJS_239_21} are the mean and the standard deviation of three DE best-fits}. For simplicity, we exclude the model parameters for the viewpoint and the background, except for the inclination. See \protect\cite{Shinn_2018_ApJS_239_21} for more details about the model.}
    \label{tbl-bestfit}
    \begin{tabular}{ccccccccccc}
        \hline
        {$\log{(L_{FUV}/L_\odot)}$} & {$\tau_{FUV}$} & {$h_d$} & {$z_d$} &  {$R_d$} & {$h_s$} & {$z_s$} & {$R_s$} &  {$\theta_{incl}$} & {Note} \\
        &  & {(kpc)} & {(kpc)} & {(kpc)} & {(kpc)} & {(kpc)} & {(kpc)} & {($^\circ$)} & \\
        {(1)} &  {(2)} & {(3)} & {(4)} &{(5)} &  {(6)} & {(7)} & {(8)} & {(9)} & {(10)} \\
        \hline
 $44.65_{ -0.23}^{ +0.31}$ & $ 2.11_{ -0.67}^{ +1.40}$ & $97.75_{-25.14}^{+28.37}$ & $ 1.38_{ -0.37}^{ +0.54}$ & $22.71_{ -0.65}^{ +0.62}$ & $19.82_{ -6.41}^{+13.54}$ & $ 0.61_{ -0.09}^{ +0.11}$ & $21.04_{ -0.28}^{ +0.30}$ & $87.39_{ -0.77}^{ +0.88}$ & this work\\

    44.230$\pm$0.022   &    0.990$\pm$0.049   &  104.903$\pm$18.309  &    2.076$\pm$0.003   &   21.890$\pm$0.170   &   28.347$\pm$1.466   &    0.768$\pm$0.003   &   21.600$\pm$0.036   &   85.907$\pm$0.182   &   \protect\cite{Shinn_2018_ApJS_239_21}  \\
        \hline
    \end{tabular}
\end{table}
\end{landscape}

\begin{figure*}
\center{
\includegraphics[scale=0.45]{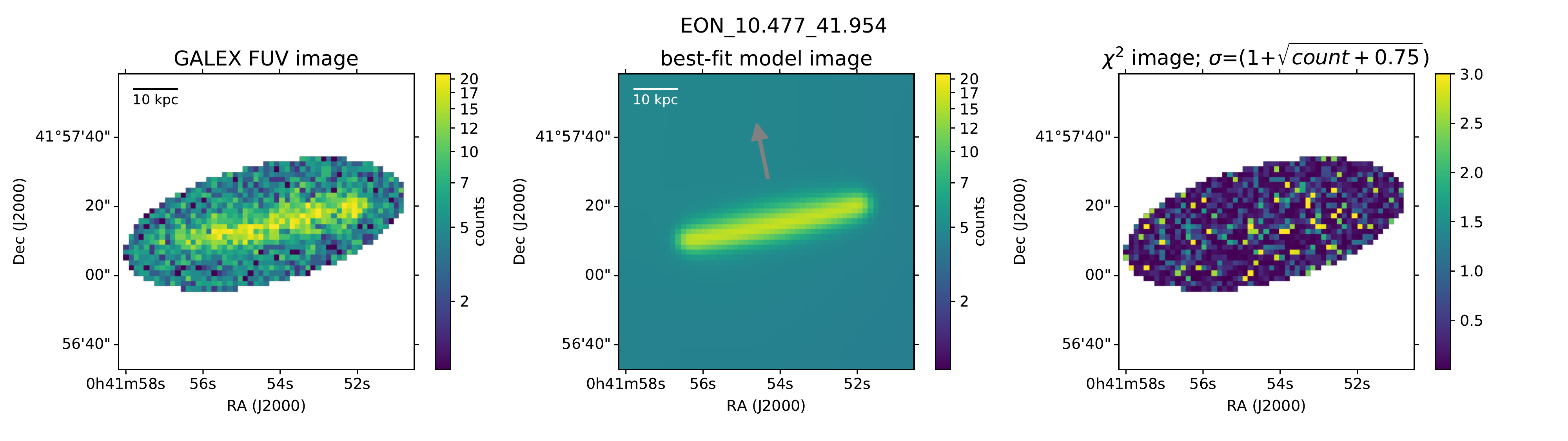}
}
\caption{Best-fit result from the Markov-Chain Monte Carlo (MCMC) analysis \nrev{on EON\_10.477\_41.954}. For the best-fit model parameters, we used the median of each PDF (see Fig.~\ref{fig-corner}). From left to right, respectively, the three panels show the \galexfull{} (\galex{}) FUV \textit{count} image, the best-fit model image, and the goodness-of-fit image. The image field-of-view is the same with the one in Fig.~\ref{fig-tg}. The color-bar maximum is set to 3 in the goodness-of-fit image, and the white areas are excluded from the MCMC analysis. The gray arrow in the best-fit model image defines the inclination angle, which increases from the arrow when moving into the page.} \label{fig-bestfit}
\end{figure*}

Fig.~\ref{fig-bestfit} shows the best-fit result found from the MCMC analysis.
We produced this image using the medians of the model-parameter PDFs (see Fig.~\ref{fig-corner}), and the goodness-of-fit image shown is the $\chi^2$ image for which the uncertainty is given by $1+\sqrt{\text{count}+0.75}$ \citep{Gehrels_1986_ApJ_303_336}, as in \cite{Shinn_2018_ApJS_239_21}, to reflect the small number of counts in the \galex{} FUV count image.
As the $\chi^2$ image shows, the model reproduces the observational image well.

\begin{figure*}
\center{
\includegraphics[scale=0.45]{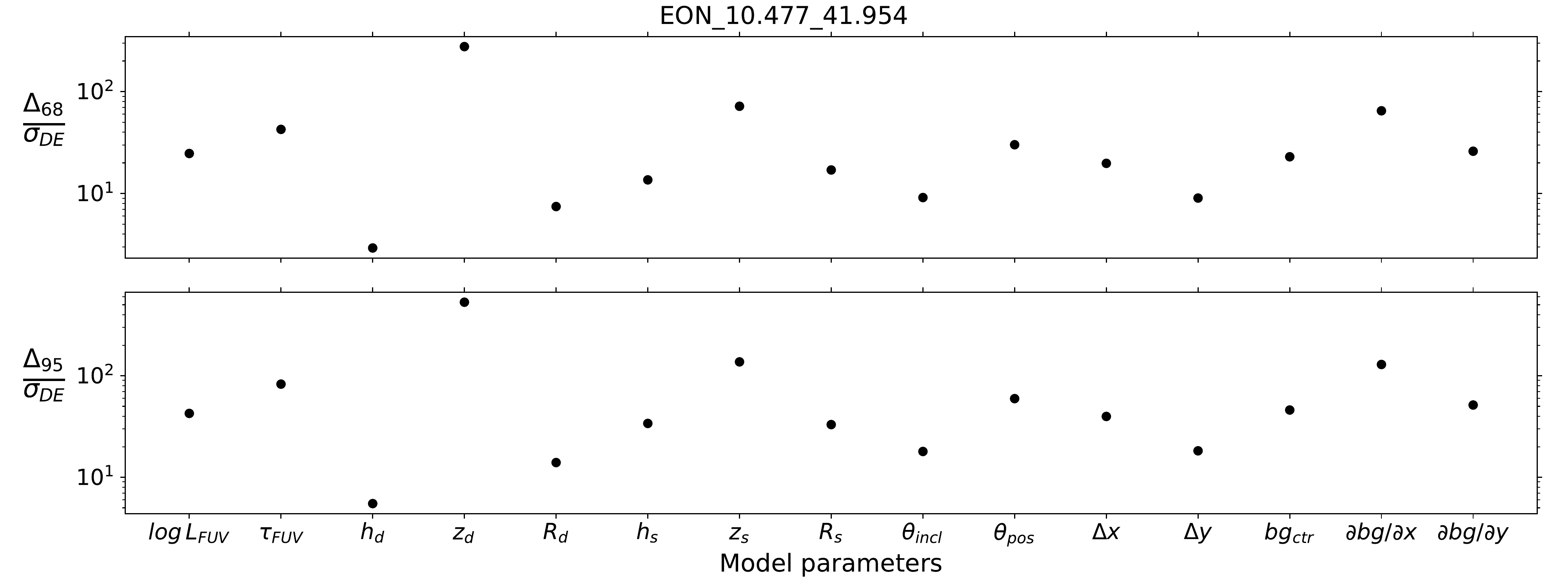}
}
\caption{Ratio of the Markov-Chain Monte Carlo (MCMC) confidence intervals $\Delta$ (this work) to the standard deviations $\sigma_{DE}$ of the DE fitting results \protect\citep{Shinn_2018_ApJS_239_21} \nrev{for EON\_10.477\_41.954}. The upper and lower panels are for the 68\% ($\Delta_{68}$) and 95\% ($\Delta_{95}$) confidence intervals, respectively.} \label{fig-err_ratio}
\end{figure*}

In our previous study \citep{Shinn_2018_ApJS_239_21}, we adopted the standard deviation of three best-fit parameters as the nominal uncertainty ($\sigma_{\text{DE}}$) for each of the best-fit model parameters, similarly as in \cite{DeGeyter_2013_A&A_550_A74,DeGeyter_2014_MNRAS_441_869} \nrev{who used five best-fits}.
To determine whether such an approach is sufficient---or whether it can at least be used as a proxy for the uncertainties---we compared the values of $\sigma_{\text{DE}}$ to the confidence intervals obtained from the MCMC analysis.
Fig.~\ref{fig-err_ratio} shows the ratios of the 68\% ($\Delta_{68}$) and 95\% ($\Delta_{95}$) confidence intervals to $\sigma_{\text{DE}}$, respectively, for all the model parameters.
The ratios are far from \nrev{unity (i.e.~perfect match) or even from} constant in either case; while the ratio is as small as $<10$ for $h_d$, it is as large as $>100$ for $z_d$.
Even excluding these two extremes ($h_d$ and $z_d$), the ratios fluctuate widely between 10 and 100.
Although such large fluctuations might be caused by the small number of tries (i.e., three), we conclude that adopting the standard deviation of several best-fit results as the uncertainty for the best-fit parameter may cause large errors in the uncertainty estimates.

\begin{figure*}
\center{
\includegraphics[scale=0.45]{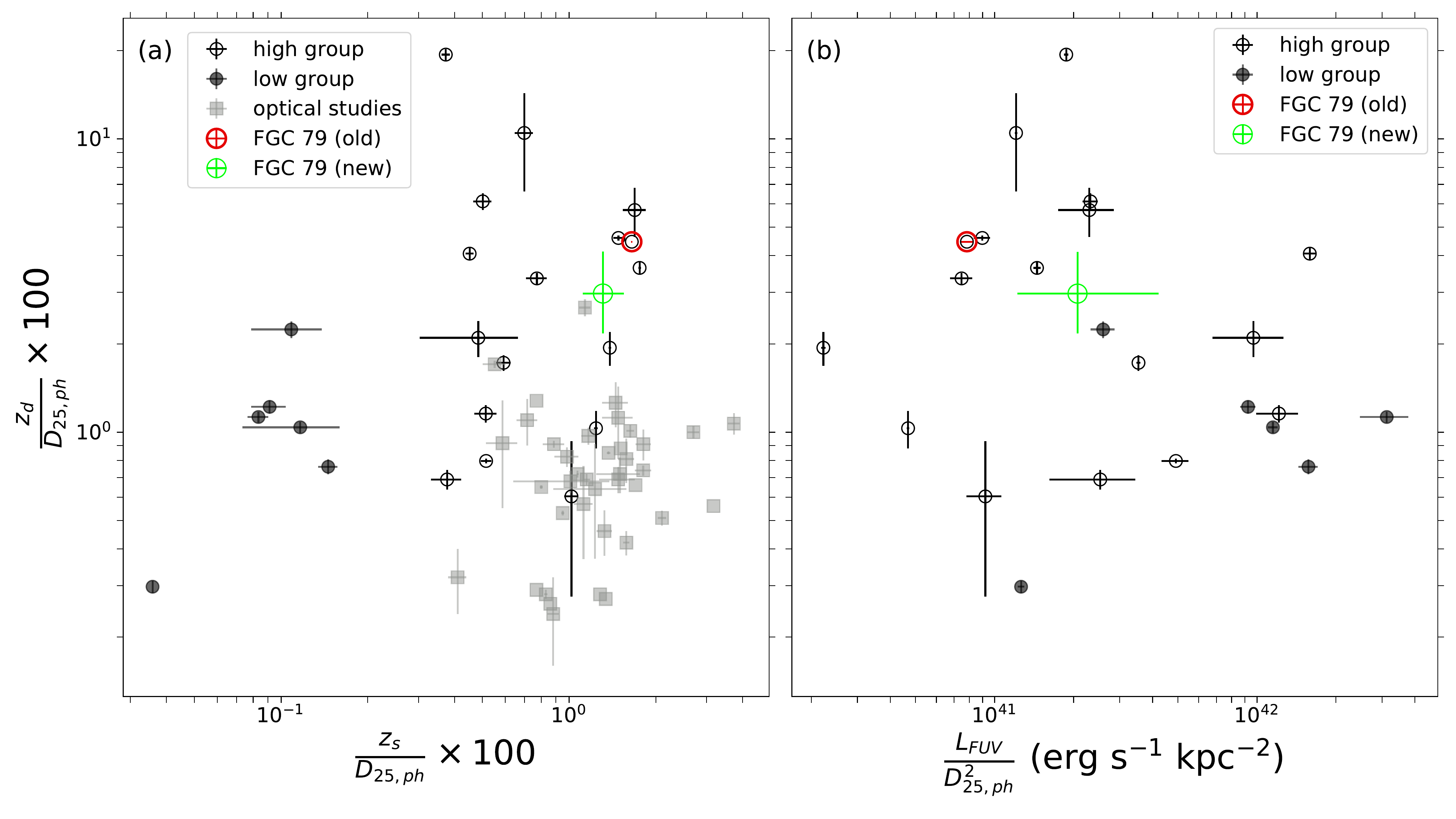}
}
\caption{Plots for size-independent galaxy comparisons. The non-colored points are adopted from \protect\cite{Shinn_2018_ApJS_239_21}. The red and green circles, respectively, represent the data point for EON\_10.477\_41.954 \nrev{(FGC 79)} from the DE analysis \citep{Shinn_2018_ApJS_239_21} and from the Markov-Chain Monte Carlo (MCMC) analysis (this work). (a) Plot showing the ratios of scale-height to galactic diameter for the dust and the light source. \textit{Open} and \textit{filled circles} represent the ``high-group'' and ``low-group,'' respectively (see text). \textit{Gray squares} represent the values obtained from optical radiative-transfer studies of other galaxies (Table 4 of \protect\citealt{Shinn_2018_ApJS_239_21}\rev{; \citealt{Peters_2017_MNRAS_464_48,Mosenkov_2018_A&A_616_A120}}). (b) Plot showing the ratio of dust scale-height to galactic diameter versus $L_{FUV}/D^2_{25,ph}$. The latter is proportional to the surface density of the FUV luminosity. The symbols are the same as in panel (a). \nrev{Both panels share the same y-axis.}} \label{fig-zD}
\end{figure*}

Using the MCMC confidence intervals, we checked how the position of EON\_10.477\_41.954 changes in the plot of $z_d/D_{25,ph}$ vs $z_s/D_{25,ph}$, which \cite{Shinn_2018_ApJS_239_21} used for target classification.
Table \ref{tbl-z_nrm} lists the relevant values based on the MCMC results, and Fig.~\ref{fig-zD}(a) shows the old (red) and new (green) position of EON\_10.477\_41.954 on the $z_d/D_{25,ph}$ vs $z_s/D_{25,ph}$ plot.
As mentioned above, the old position falls within the 95\% confidence interval of the new position.
Both $z_s/D_{25,ph}$ and $z_d/D_{25,ph}$ decrease a little ($\sim20-30$\%), but EON\_10.477\_41.954 still falls in the ``high-group'' region ($z_s/D_{25,ph}>0.2$).
The ``high-group'' and ``low-group'' include galaxies, respectively with and without, substantial extraplanar dust, as classified by \cite{Shinn_2018_ApJS_239_21}.
In comparison to the values obtained from the optical radiative-transfer studies (gray squares) that represent galactic think disks\del{ (see Table 4 of }\delref{\citealt{Shinn_2018_ApJS_239_21})}, EON\_10.477\_41.954 has a similar value of $z_s/D_{25,ph}$ but a higher value of $z_d/D_{25,ph}$.
\rev{The optical-study data points are from Table 4 of \cite{Shinn_2018_ApJS_239_21}, \cite{Peters_2017_MNRAS_464_48}, and \cite{Mosenkov_2018_A&A_616_A120}. When plotting the points of \cite{Peters_2017_MNRAS_464_48}, we used the distances listed in \cite{Peters_2017_MNRAS_464_2} and the major axis diameter ($D_{25}$) from NASA/IPAC Extragalactic Database (\url{https://ned.ipac.caltech.edu/}). Also, we only included the fast-rotating (maximum circular velocity $\upsilon_{\text{max}}>130$ \kms) galaxies in order to be consistent with other optical studies (see \citealt{Peters_2017_MNRAS_464_48}).}
In Fig.~\ref{fig-zD}(b), we show for reference the corresponding plot of $z_d/D_{25,ph}$ vs. $L_{FUV}/D^2_{25,ph}$, as in \cite{Shinn_2018_ApJS_239_21}.
The value of $L_{FUV}/D^2_{25,ph}$ is increased, but the old value is within the 95\% confidence interval of the new value.

\begin{table}
    \centering
    \caption{Ratio of Scale-height to Galactic Diameter \nrev{for EON\_10.477\_41.954}. \rev{The results of \protect\cite{Shinn_2018_ApJS_239_21} are appended at the last row for comparison.} Column (1) is from \protect\cite{Shinn_2018_ApJS_239_21}, and Columns (2)-(3) are calculated from $z_s$ and $z_d$ from Table \ref{tbl-bestfit} using Column (1). \rev{Column (4) is a note for the row.}}
    \label{tbl-z_nrm}
    \begin{tabular}{cccc}
        \hline
        {$D_{25,ph}$} &  {{\large$\frac{z_{s}}{D_{25,ph}}$}$\times100$} & {{\large$\frac{z_{d}}{D_{25,ph}}$}$\times100$} & {Note} \\
        {(kpc)} & & \\
        {(1)} &  {(2)} & {(3)} & {(4)} \\
        \hline
                      46.5 & $ 1.31_{ -0.19}^{ +0.24}$ & $ 2.97_{ -0.80}^{ +1.16}$ & this work\\

       46.5 &      1.650$\pm$0.006      &      4.461$\pm$0.006      &    \protect\cite{Shinn_2018_ApJS_239_21} \\

        \hline
    \end{tabular}
\end{table}

\ndel{Since our main concern is the target classification using two scale-heights $z_s$ and $z_d$, it is worth checking how their confidence intervals are determined, even based on the poorly-resolved \galex{} image of EON\_10.477\_41.954 (Fig.~\ref{fig-bestfit}). Thus, we made a reference model image and see the difference between it and the best-fit model image (see Fig.~\ref{fig-diff}). The reference image is produced with the best-fit model parameters except two parameters $z_s$ and $z_d$, where the values are given by their best-fit values plus 1-$\sigma$ deviations, i.e. $z_s=0.72$ kpc and $z_d=1.92$ kpc (see Table \ref{tbl-bestfit}). As seen from Fig.~\ref{fig-diff}, the reference image is greater than the best-fit image along the galactic disc by a nearly-fixed amount. This excess is reasonable because the reference image is produced with higher $z_s$ and $z_d$ which allow the starlight to escape the galaxy more easily. Fig.~\ref{fig-diff} demonstrates that $z_s$ and $z_d$ can be well constrained even when the galaxy image is poorly-resolved, thanks to the pixel-value differences of the model images.}

\subsection{\nrev{Test on the Validity of the MCMC Method Applied to the Poorly-resolved Galaxy Image}} \label{ana-res-test}
\begin{figure*}
\center{
\includegraphics[scale=0.45]{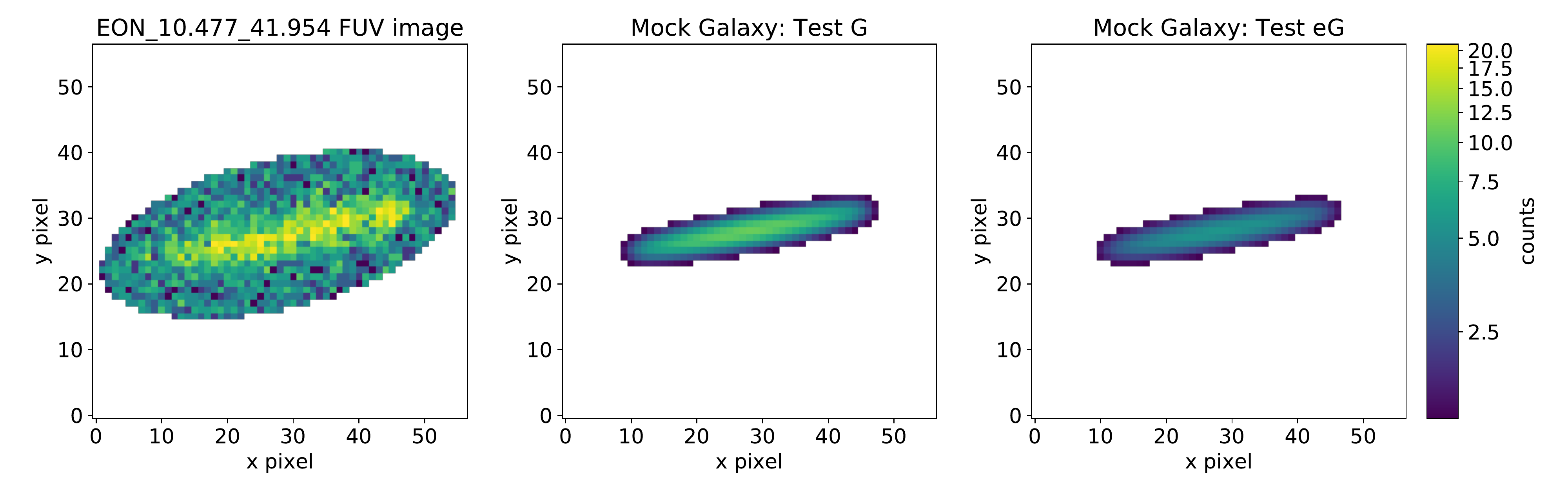}
}
\caption{\nrev{Comparison between the real galaxy image and the mock galaxy images. From left to right, respectively, the three panels show the \galexfull{} (\galex{}) FUV \textit{count} image of EON\_10.477\_41.954, the Test G mock galaxy image, and the Test eG mock galaxy image (see Table \ref{tbl-mock}). The white areas are excluded from the MCMC analysis.}} \label{fig-mock}
\end{figure*}

\nrev{
EON\_10.477\_41.954 is poorly-resolved in the \galex{} FUV image (Fig.~\ref{fig-tg}), hence it might be hard to infer the true vertical structure perpendicular to the galactic plane.
That is, the model parameters determined from the MCMC analysis can be different from the true values.
In order to see how the low spatial resolution affects the parameter estimation of the MCMC analysis, we made two mock galaxies and performed the MCMC analysis in the same way we applied to EON\_10.477\_41.954.
Table \ref{tbl-mock} lists the input parameters for the two mock galaxies (Test G and eG).
We adopted the same distance with EON\_10.477\_41.954 (160 Mpc, \citealt{Shinn_2018_ApJS_239_21}), the same pixel scale and point-spread-function with the \galex{} FUV band, for the mock images to be similar to the FUV image of EON\_10.477\_41.954.
The background was set to zero for simplicity.
Fig.~\ref{fig-mock} shows the two mock galaxy images in addition to the \galex{} FUV image of EON\_10.477\_41.954, for comparison.
}

\begin{landscape}
\begin{table}
    \caption{\nrev{Summary of the Tests on Two Mock Galaxies. Column (1), the name of the mock galaxy and the relevant information ; Column (2), the galaxy luminosity in the FUV band; Column (3), the optical depth in the FUV band through the galactic center when viewed face-on; Column (4), the scale-length of the dust component; Column (5), the scale-height of the dust component; Column (6), the cutoff radius of the dust component; Column (7), the scale-length of the light-source component; Column (8), the scale-height of the light-source component; Column (9), the cutoff radius of the light-source component; Column (10), the inclination angle; Column (11), the position angle; Column (12), the shift in the horizontal direction on the image plane; and Column (13), the shift in the vertical direction on the image plane.}}
    \label{tbl-mock}
    \begin{tabular}{cccccccccccccc}
        \hline
        {Mock Galaxy} & {$\log{(L_{FUV}/L_\odot)}$} & {$\tau_{FUV}$} & {$h_d$} & {$z_d$} &  {$R_d$} & {$h_s$} & {$z_s$} & {$R_s$} &  {$\theta_{incl}$} & {$\theta_{pos}$} & {$\Delta x$} & {$\Delta y$} \\
        {} &  &  & {(kpc)} & {(kpc)} & {(kpc)} & {(kpc)} & {(kpc)} & {(kpc)} & {($^\circ$)} & {($^\circ$)} & {(pixels)} & {(pixels)} \\
        {(1)} &  {(2)} & {(3)} & {(4)} &{(5)} &  {(6)} & {(7)} & {(8)} & {(9)} & {(10)} & {(11)} & {(12)} & {(13)} \\
        \hline
        Test G\\
        \hline
        model input & 
            48.7 &             2.1 &            13.2 &             0.5 &            26.0 &             8.8 &             0.2 &            22.0 &            89.0 &           100.0 &             0.0 &             0.0\\
        MCMC output & 
 $49.21_{ -0.24}^{ +0.30}$ & $ 4.27_{ -1.41}^{ +1.76}$ & $10.87_{ -1.95}^{ +2.53}$ & $ 0.35_{ -0.07}^{ +0.10}$ & $26.08_{ -1.28}^{ +1.36}$ & $ 7.26_{ -1.57}^{ +2.25}$ & $ 0.06_{ -0.03}^{ +0.06}$ & $22.23_{ -0.45}^{ +0.50}$ & $89.61_{ -1.15}^{ +1.33}$ & $99.98_{ -0.24}^{ +0.26}$ & $-0.01_{ -0.21}^{ +0.22}$ & $-0.01_{ -0.08}^{ +0.08}$\\

        \hline
        \hline
        Test eG \\
        \hline
        model input & 
            48.7 &             2.1 &            13.2 &             0.5 &            26.0 &             8.8 &             0.2 &            22.0 &            89.0 &           100.0 &             0.0 &             0.0\\
&              - &             2.1 &            13.2 &             1.5 &            26.0 &               - &               - &               - &               - &               - &               - &               -\\
        MCMC output & 
 $48.95_{ -0.27}^{ +0.30}$ & $ 4.82_{ -1.47}^{ +1.68}$ & $14.58_{ -3.13}^{ +3.90}$ & $ 0.68_{ -0.24}^{ +0.32}$ & $26.08_{ -1.56}^{ +1.44}$ & $ 8.55_{ -2.44}^{ +3.68}$ & $ 0.11_{ -0.06}^{ +0.14}$ & $22.26_{ -0.70}^{ +0.85}$ & $89.92_{ -1.18}^{ +1.28}$ & $100.01_{ -0.46}^{ +0.45}$ & $ 0.02_{ -0.34}^{ +0.38}$ & $-0.02_{ -0.11}^{ +0.11}$\\

        \hline
    \end{tabular}
\end{table}
\end{landscape}

\nrev{
Test G represents a spiral galaxy without substantial extraplanar dust.
First, we chose $R_s$ for the angular diameter to be comparable to that of EON\_10.477\_41.954.
Then, we set $z_d$ to be comparable to the ratio $z_d/D_{25,ph}\sim0.01$ of the optical studies (Fig.~\ref{fig-zD}), while we set $z_s$ to be smaller than the ratio $z_s/D_{25,ph}\sim0.01$ of the optical studies (Fig.~\ref{fig-zD}), since the \galex{} FUV image traces the young stellar population which has a lower scale-height than the old stellar population \citep[see][]{Bahcall_1980_ApJS_44_73,Wainscoat_1992_ApJS_83_111,Martig_2014_MNRAS_442_2474}.
The values of $\tau_{FUV}$, $h_d$, and $h_s$ were set referring the values used to produce a test image in \cite{DeGeyter_2013_A&A_550_A74}.
We set $\tau_{FUV}$ to be comparable to the $V$-band face-on optical depth of \cite{DeGeyter_2013_A&A_550_A74}.
The values of $h_d$ and $h_s$ were set to be double of \citeauthor{DeGeyter_2013_A&A_550_A74}'s, in order to make the brightness along the galactic plane more uniform like EON\_10.477\_41.954.
We set $R_d$ to be larger than $R_s$ to prevent the light sources from being in the dust-free region.
$L_{FUV}$ was chosen for the mock galaxy to have similar counts per pixel as the \galex{} image of EON\_10.477\_41.954.
When converting the count-per-sec image to the count image, we used a fake relative response image whose value is unity at all pixels, for simplicity; hence $L_{FUV}$ is unrealistic.
Test eG represents a spiral galaxy with substantial extraplanar dust.
We added a second dust disk to Test G.
The second dust disk shares the same parameters with that of Test G, except $z_d$ which is three times higher.
}

\begin{figure*}
\center{
\includegraphics[scale=0.45]{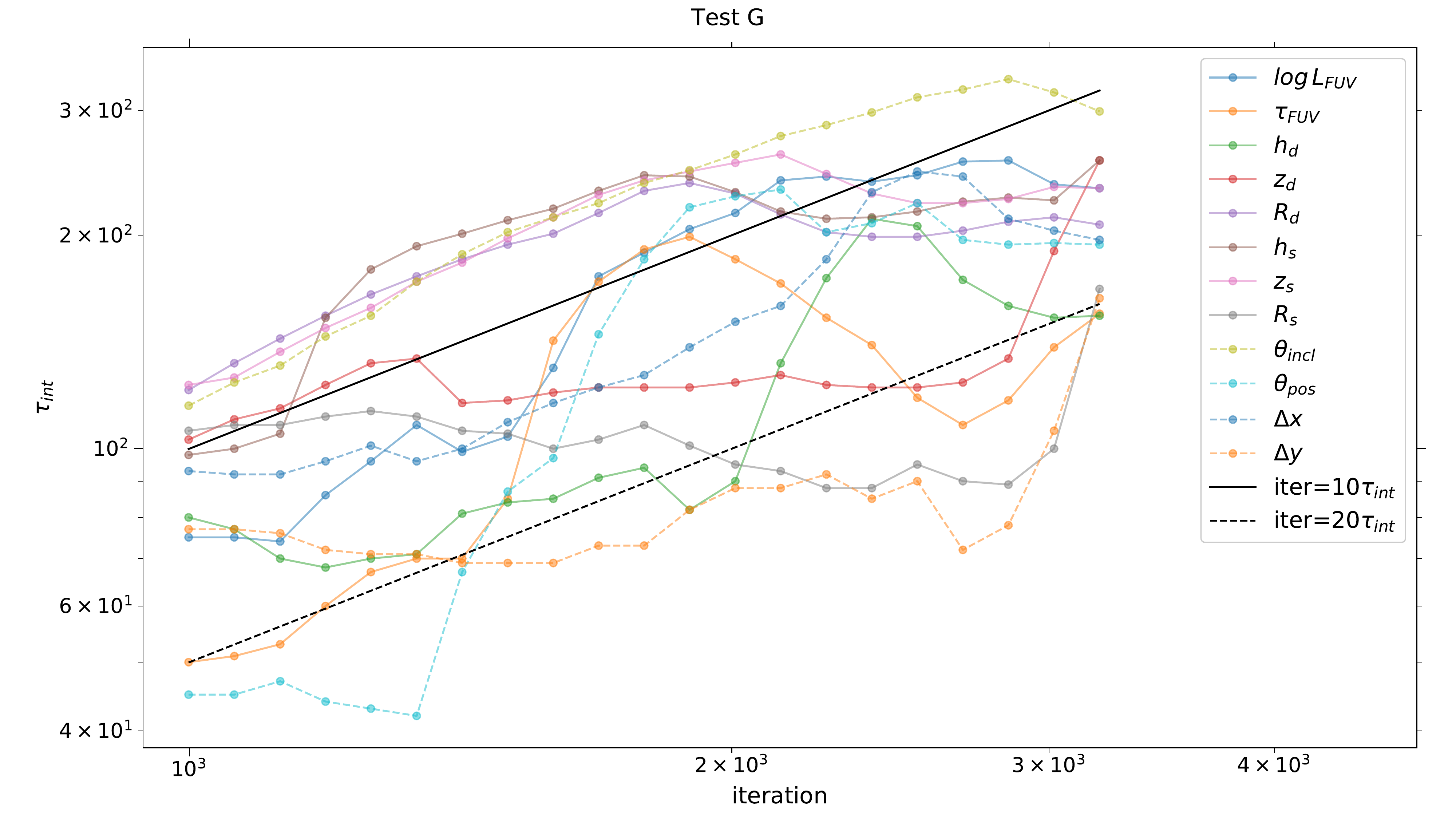}
}
\caption{\nrev{Evolution of the integrated autocorrelation time $\tau_{int}$ for Test G. Each line corresponds to the model parameter listed in the legend. The black solid and dashed lines indicate the loci of $\mathrm{iteration}=10\times\tau_{int}$ and $\mathrm{iteration}=20\times\tau_{int}$, respectively.}} \label{fig-iat-G}
\end{figure*}

\begin{figure*}
\center{
\includegraphics[scale=0.45]{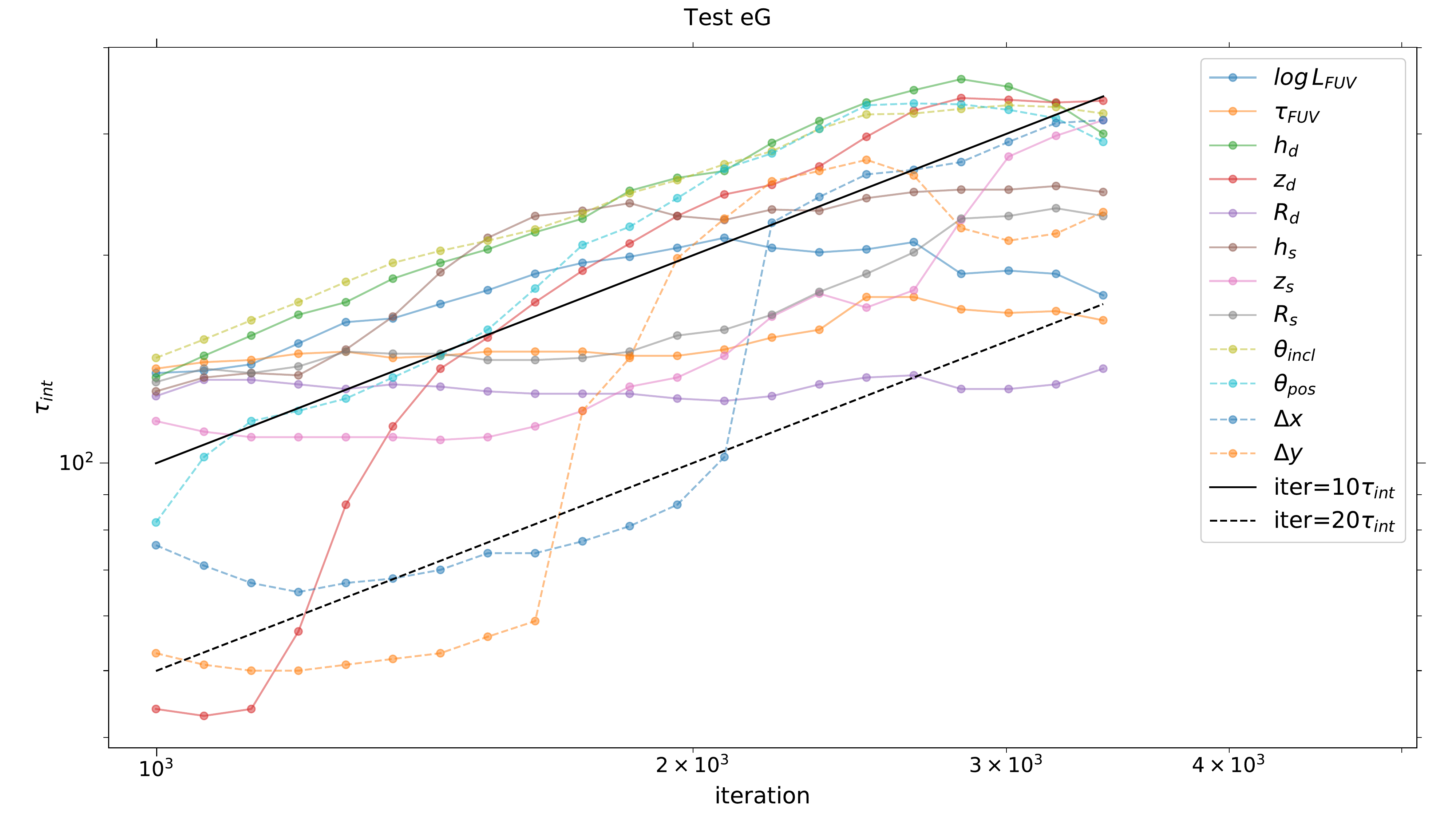}
}
\caption{\nrev{Evolution of the integrated autocorrelation time $\tau_{int}$ for Test eG. Each line corresponds to the model parameter listed in the legend. The black solid and dashed lines indicate the loci of $\mathrm{iteration}=10\times\tau_{int}$ and $\mathrm{iteration}=20\times\tau_{int}$, respectively.}} \label{fig-iat-eG}
\end{figure*}

\nrev{
We performed the MCMC analysis on the two mock galaxies (Test G and eG) in the same way as on EON\_10.477\_41.954 (see section \ref{ana-res-mcmc} and \ref{ana-res-cnvg}).
First, we found the maximum likelihood region using the DE method, and then used the final parameter vectors from the DE analysis as the initial MCMC walkers.
The control variables of the DE method are the same with those for Test A-C of \cite{Shinn_2018_ApJS_239_21}; therefore, the number of the initial MCMC walkers is 90.
For both Test G and eG, the burn-in stage of 5000 iterations were needed.
We stopped the MCMC sampling when $\tau_{int}$ of all the model parameters fall between the two lines of $\mathrm{iteration} = 10\,\tau_{int}$ and $\mathrm{iteration} = 20\,\tau_{int}$ (see Fig.~\ref{fig-iat-G} and \ref{fig-iat-eG}), where $\tau_{int}$ is calculated from the mean of the 90 walkers, $X_i=\frac{1}{90}\sum^{90}_{i=1} x_i$.
We could not push further iterations because the repeating emergence of correlated chains makes $\tau_{int}$ increase abruptly (see section \ref{ana-res-cnvg}).
These two lines corresponds to the ESS of $10\times90=900$ and $20\times90=1800$, respectively.
These ESSs are close to the numbers of independent samples required to determine the 0.025 quantile to within $\pm0.01$ and $\pm0.0075$, respectively, with probability 0.95; i.e., 936 and 1665 \citep{Raftery_1992_inproc}.
}\nrrev{The achieved ESSs for Test G and eG ($\sim900-1800$) are smaller than those for EON\_10.477\_41.954 ($>5,650$), because the correlated MCMC chains emerge more frequently for Test G and eG. We guess this might be related to the fact that the mock images are produced by the model itself which is used for the MCMC sampling.}

\begin{figure*}
\center{
\includegraphics[scale=0.25]{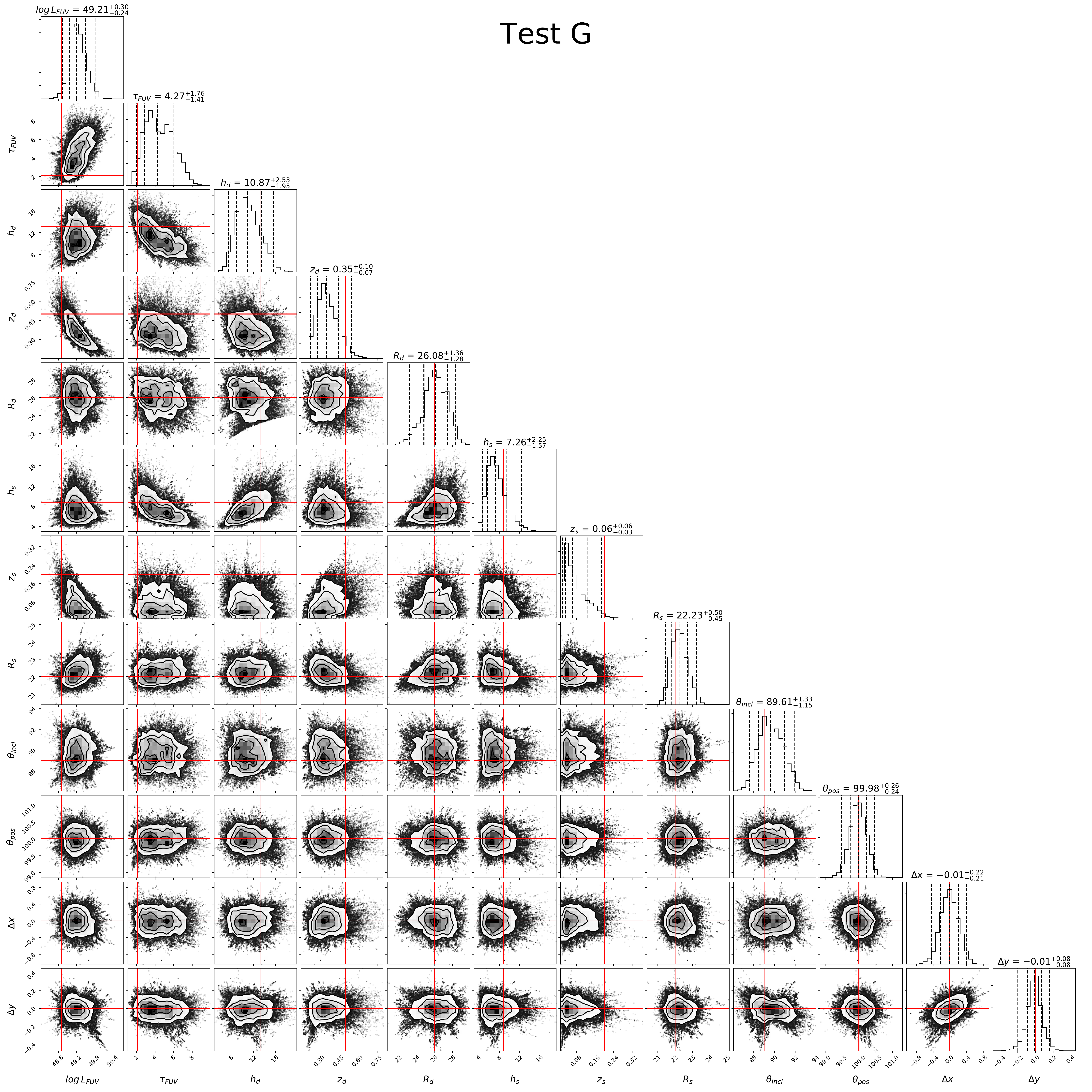}
}
\caption{\nrev{Corner plot for Test G. The diagonal panels show the marginal probability distribution functions of the model parameters, where the five vertical dashed lines represent the median, 68\%, and 95 \% confidence intervals. The off-diagonal panels show the pairwise relations, with no smoothing applied. The red lines indicate the model input values (see Table \ref{tbl-mock}). The lower end of the $z_s$ distribution is limited by the parameter space limit, 0.01 kpc. (See the separate sheet for a larger version of this figure.)}} \label{fig-corner-G}
\end{figure*}

\begin{figure*}
\center{
\includegraphics[scale=0.25]{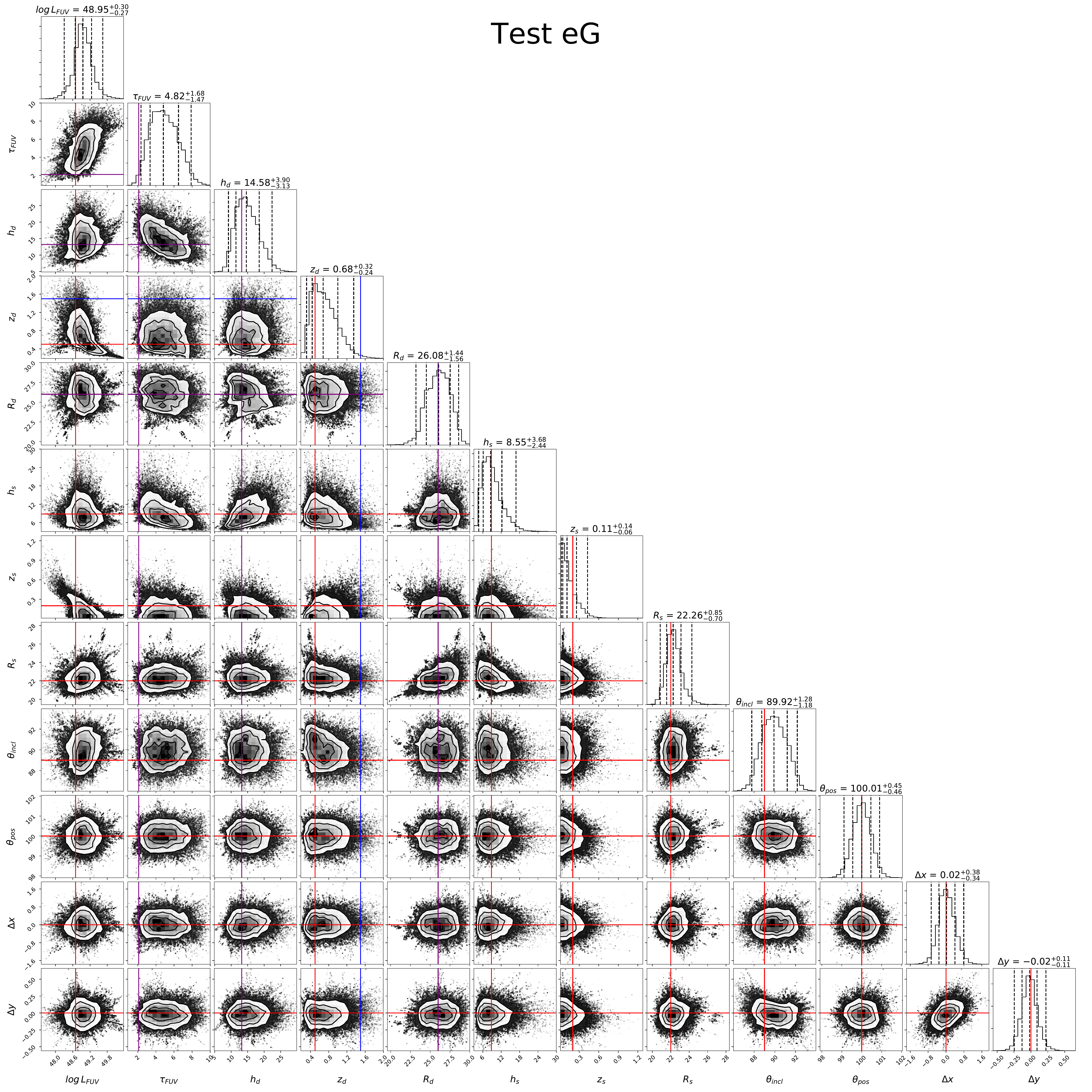}
}
\caption{\nrev{Corner plot for Test eG. The diagonal panels show the marginal probability distribution functions of the model parameters, where the five vertical dashed lines represent the median, 68\%, and 95 \% confidence intervals. The off-diagonal panels show the pairwise relations, with no smoothing applied. The red and blue lines indicate the model input values (see Table \ref{tbl-mock}). The lower end of the $z_s$ distribution is limited by the parameter space limit, 0.01 kpc. (See the separate sheet for a larger version of this figure.) }} \label{fig-corner-eG}
\end{figure*}

\nrev{
Fig.~\ref{fig-corner-G} and \ref{fig-corner-eG} show the corner plots for Test G and eG, and the determined model parameters are listed in Table \ref{tbl-mock}.
\nrrev{In both figures, the parameter $z_s$ is limited at its lower end by the parameter space limit (0.01 kpc) that was set to be small enough considering the expected $z_s$ for young stellar populations. Also, all parameters show single-peaked marginal distributions in both tests, except $\tau_{FUV}$ of Test G (Fig.~\ref{fig-corner-G}) which is weakly double-peaked.}
As Fig.~\ref{fig-corner-G} shows, for Test G, all the input parameters are recovered within 2-$\sigma$ (95\%) range.
However, some parameters show larger deviations from the input values than the others.
$L_{FUV}$, $\tau_{FUV}$, $z_d$ and $z_s$ show a nearly 2-$\sigma$ deviation, while $R_d$, $\theta_{pos}$, $\Delta x$ and $\Delta y$ fall right on the input values.
We note that the former (especially $\tau_{FUV}$, $z_d$ and $z_s$) is the parameters related to the galactic vertical structure, while the latter is not.
Based on this contrast, we suppose that the larger deviations of $\tau_{FUV}$, $z_d$ and $z_s$ are likely due to the poor spatial resolution of the mock galaxy image, which distorts the likelihood surface for the maximum point to be deviated.
The same kind of deviation would also be expected for the MCMC results of EON\_10.477\_41.954, since it is poorly-resolved in the \galex{} FUV image (Fig.~\ref{fig-tg}).
We here note that $z_d$ and $z_s$, our main interest for the target classification, are underestimated than the input values.
In Fig.~\ref{fig-corner-eG} for Test eG, we can see that the existence of substantial extraplanar dust makes $z_d$ and $z_s$ overestimated.
These two parameters are greater than those of Test G (see Table \ref{tbl-mock}): (Test eG : Test G) = ($ 0.68_{ -0.24}^{ +0.32}$ : $ 0.35_{ -0.07}^{ +0.10}$) kpc for $z_d$ = ($ 0.11_{ -0.06}^{ +0.14}$ : $ 0.06_{ -0.03}^{ +0.06}$) kpc for $z_s$.
Therefore, we think that our methodology for finding the edge-on galaxies with substantial extraplanar dust on Fig.~\ref{fig-zD}(a) would work even for the poorly-resolved edge-on galaxies.
}

\section{Discussion \label{discu}}
We have carried out an MCMC analysis of the modeling of the \galex{} FUV image of EON\_10.477\_41.954 in order to determine the uncertainties of the model parameters more accurately.
Since the MCMC analysis is basically a sampling of the likelihood surface, it was necessary to measure the convergence of the sampling.
We monitored the convergence using the ESS ($N/\tau_{int}$, see section \ref{ana-res-cnvg}), and Fig.~\ref{fig-iat} shows the evolution of $\tau_{int}$ over the iterations.
In section \ref{ana-res-cnvg}, we noted that it was necessary to stop and resume the MCMC run several times, since $\tau_{int}$ increased abruptly at some points and hence the ESS did not increase.
This demonstrates that running an MCMC analysis without monitoring $\tau_{int}$ is meaningless---or at best problematic---because $N$ can sometimes increase without increasing the ESS.
We therefore conclude that an MCMC analysis should always be accompanied by proper convergence monitoring.

In Fig.~\ref{fig-conf_evol}, we showed the evolution of the confidence intervals during the iterations.
The 68\% and 95\% confidence intervals stabilized in position and width after all the parameters achieved almost constant values of $\tau_{int}$, say, for iterations $\ga20,000$.
These intervals are unstable at early iterations, especially for iterations $\la5,000$, where the values of $\tau_{int}$ are not constant; again, this clearly shows the importance of convergence monitoring.
In Fig.~\ref{fig-conf_evol}, we also showed the best-fit results from \cite{Shinn_2018_ApJS_239_21} for comparison, labeling them as ``DE best.''
These ``DE best'' points are not on the modes of the PDFs (see also Fig.~\ref{fig-corner}), perhaps due to the nature of DE algorithm.
The DE algorithm is a sort of global-optimization method, and it seeks the maximum value of the likelihood surface by mixing several model-parameter vectors.
When all the model-parameter vectors converge to a certain value, this approach concludes that the maximum value has been found.
Therefore, if all the model-parameter vectors are very close each other, but are evolving very slowly in a clump, the DE method cannot pinpoint the maximum value, instead ending up with some value close to the maximum.

In our previous study \citep{Shinn_2018_ApJS_239_21}, we adopted $\sigma_{\text{DE}}$ as the nominal uncertainty for the model parameters, similarly as in \cite{DeGeyter_2013_A&A_550_A74,DeGeyter_2014_MNRAS_441_869}.
To check the validity of this approach, we plotted the ratios of $\sigma_{\text{DE}}$ to $\Delta_{68}$ and $\Delta_{95}$, respectively.
As shown in Fig.~\ref{fig-err_ratio}, the ratios are neither unity nor even constant; instead, they range from $\ga5$ up to $>100$.
This means that $\sigma_{\text{DE}}$ is not a suitable proxy for the model-parameter uncertainty.
The values of $\sigma_{\text{DE}}$ were obtained from several best-fit values that were found by a global-optimization method.
There is no reason for $\sigma_{\text{DE}}$ to reflect the shape of the likelihood surface around the maximum, because the global-optimization method only finds that maximum point, regardless of the shape of the likelihood surface around it.
This may be the cause of the large fluctuations in ${\Delta_{68}}/{\sigma_{\text{DE}}}$ and ${\Delta_{95}}/{\sigma_{\text{DE}}}$.
The difficulty in pinpointing the exact maximum that the DE method can experience, which is mentioned in the previous paragraph, could also make the fluctuations of ${\Delta_{68}}/{\sigma_{\text{DE}}}$ and ${\Delta_{95}}/{\sigma_{\text{DE}}}$ worse.

From our MCMC analysis results, we can now position EON\_10.477\_41.954 more accurately in the target-classification plot [Fig.~\ref{fig-zD}(a)].
The new position (green) has lower values than the old (red) position on both axes ($\sim20-30$\% decrease), but it nevertheless includes the old position within the 95\% confidence intervals.
This shows that EON\_10.477\_41.954 still belongs to the ``high-group'' ($z_s/D_{25,ph}>0.2$, see \citealt{Shinn_2018_ApJS_239_21}), even with the new position.
This membership seems more likely by comparison to the positions of galactic thin disks (gray squares) as determined from optical radiative studies.
The relative positions are still the same as in \cite{Shinn_2018_ApJS_239_21}; the new position of EON\_10.477\_41.954 has a similar value of $z_s/D_{25,ph}$ but a higher value of $z_d/D_{25,ph}$ than the galactic thin disks.
These values of $z_s/D_{25,ph}$ indicate\del{s} that the young stellar population (which is bright in the ultraviolet) has a similar scale-height to the old stellar population (which is bright in the optical).
This is not consistent with the general picture of galactic structure---a young stellar population usually has a lower scale-height than an old stellar population \citep[cf.,][]{Bahcall_1980_ApJS_44_73,Wainscoat_1992_ApJS_83_111,Martig_2014_MNRAS_442_2474}.
The existence of substantial extraplanar dust probably causes this inconsistency---and also the higher value of $z_d/D_{25,ph}$---\rev{since the extraplanar dust scatters the starlight further away from the galactic plane,} as explained in \cite{Shinn_2018_ApJS_239_21}.

\nrev{The membership to ``high-group'' becomes more solid when we consider the $\sim$2-$\sigma$ underestimations of $z_d$ and $z_s$ due to the poor spatial resolution of the \galex{} image (see section \ref{ana-res-test}).}
\nrrev{In the test with mock galaxies in section \ref{ana-res-test}, we did not add any synthetic noise (e.g., instrumental and Poisson noises) to the mock images. These noises are random at each pixel and do not change the shape of the mock galaxy systematically, like lowering the spatial resolution of the galaxy image, which caused the underestimations of $z_d$ and $z_s$. Therefore, we think the addition of those synthetic noises would not shift the mode of the model parameter distributions, but rather broaden the confidence intervals of the parameters. In this sense, the underestimation of $z_d$ and $z_s$ could be weaker than $\sim$2-$\sigma$ due to the broadening of the confidence intervals, if the noises are added to the mock galaxy.}

It is hard to guess how other galaxies may move in the target-classification plot [Fig.~\ref{fig-zD}(a)] after an MCMC analysis, since each of them has a different likelihood surface around the maximum point.
If the degree of movement is not much different from the case of EON\_10.477\_41.954 (i.e., $\sim20-30$\%), we speculate that the target-classification of \cite{Shinn_2018_ApJS_239_21} will remain unchanged.
However, we cannot be sure of this until an MCMC analysis has been carried out for each target\nrev{, which is in preparation}.
If we assume a similar amount of movement in the plot of $z_d/D_{25,ph}$ vs. $L_{FUV}/D^2_{25,ph}$ [Fig.~\ref{fig-zD}(b)], the conclusion of \cite{Shinn_2018_ApJS_239_21} that $L_{FUV}/D^2_{25,ph}$ is not a good discriminator for galaxies with substantial extraplanar dust would also remain unchanged, since the ``high-group'' and ``low-group'' are mixed along the $L_{FUV}/D^2_{25,ph}$ axis.

\section{Conclusions \label{concl}}
We have revisited the target EON\_10.477\_41.954, which \cite{Shinn_2018_ApJS_239_21} classified as a galaxy with substantial extraplanar dust (``high-group''), in order to obtain a more accurate determination of the uncertainties in the model parameters and hence a more rigorous target classification.
In our previous study \citep{Shinn_2018_ApJS_239_21}, we used the standard deviation of three best-fit results ($\sigma_{\text{DE}}$) as the nominal uncertainty for each model parameter.
Here, we instead performed an MCMC analysis of the likelihood surface and obtained more accurate uncertainties for fifteen model parameters.
We monitored the convergence of the MCMC sampling\rev{---which is usually neglected in the literature but should not be} \nrev{to measure the reliability of the }\nddel{sample}\nrrev{sampling}\rev{---}using $\tau_{int}$, and we achieved an ESS of $>5,650$ for all the model parameters.
We had to stop and resume the MCMC sampling several times, since $\tau_{int}$ increased abruptly at some iterations, and hence the ESS did not increase.
We demonstrated that the confidence intervals are stable when the values of $\tau_{int}$ are almost constant, but they are unstable when those values are increasing at the beginning of the iterations.
\nrrev{Additionally, in order to see how the shape of the likelihood surface is affected by the low spatial-resolution of the EON\_10.477\_41.954 \galex{} FUV image, we carried out two tests with mock galaxy images. We found that some model parameters could be deviated from the true value up to $\sim$2-$\sigma$ level---e.g., $z_d$ and $z_s$ are underestimated by $\sim$2-$\sigma$---but our methodology to find the galaxies with substantial extraplanar dust using $z_d$ and $z_s$ (see Fig.~\ref{fig-zD}) still works.}

The modes of the parameter PDFs are different from the best-fit parameters found by the global-optimization DE method in \cite{Shinn_2018_ApJS_239_21}.
The slow approach of the DE method to the maximum-likelihood point seems to cause this difference.
In order to check if $\sigma_{\text{DE}}$ can be used as a proxy for the model-parameter uncertainty, we compared the confidence intervals ($\Delta_{68}$ and $\Delta_{95}$) to $\sigma_{\text{DE}}$. We obtained the negative result; that ${\Delta_{68}}/{\sigma_{\text{DE}}}$ and ${\Delta_{95}}/{\sigma_{\text{DE}}}$ are \nddel{far from unity and are not even constant}\nrrev{neither unity nor constant, but rather scattered over $\sim5-100$}.
The new position of EON\_10.477\_41.954 in the target-classification plot ($z_d/D_{25,ph}$ vs. $z_s/D_{25,ph}$) decreases by about $20-30$\% in both axes, but the old position is included within the 95\% confidence interval of the new one.
EON\_10.477\_41.954 still belongs to the ``high-group,'' even judging from the new position; $z_s/D_{25,ph}$ is $>0.2$, and its position relative to the positions of galactic thin disks obtained from optical radiative-transfer studies is still the same as in \cite{Shinn_2018_ApJS_239_21}.
\nrev{The situation is better when we consider the $\sim$2-$\sigma$ underestimations of $z_d$ and $z_s$ due to the poor spatial resolution of the \galex{} image.}
The changes in the positions of other target galaxies in the target-classification plot is hard to predict; however, if the amount of change is similar to the case of EON\_10.477\_41.954, the target classification would remain unchanged.

\section*{Acknowledgements}
J.-H.S. is grateful to the anonymous referee for his-or-her useful comments that make the manuscript clearer and to Kwang-Il Seon for providing the galaxy model.




\bibliographystyle{mnras}
\bibliography{extraplanar_dust} 







\bsp	
\label{lastpage}
\end{document}